\documentclass[twocolumn,prc,aps,showpacs,superscriptaddress, nofootinbib,10pt]{revtex4-1}
\usepackage{graphicx,graphics}
\usepackage{amsmath}
\usepackage{amssymb}
\usepackage{epstopdf}
\usepackage{amstext}
\usepackage{amsthm}
\usepackage{amsfonts}
\usepackage{latexsym}
\usepackage{array}
\usepackage{xfrac}
\usepackage{color}
\usepackage{subfigure}
\usepackage{multirow}
\usepackage{fontenc}
\usepackage{bm}
\usepackage{adjustbox}
\usepackage{dcolumn}
\usepackage{soul}
\usepackage{hyperref}
\hypersetup{colorlinks=true,breaklinks,linkcolor=red,citecolor=blue}
\usepackage{color}

\begin{document}

\title{Study of the core-crust transition in neutron stars with finite-range 
interactions: the dynamical method}

\author{C. Gonzalez-Boquera}
\email{cgonzalezboquera@ub.edu}
\affiliation{Departament de F\'isica Qu\`antica i Astrof\'isica and Institut de Ci\`encies del Cosmos (ICCUB), 
Facultat de F\'isica, Universitat de Barcelona, Mart\'i i Franqu\`es 1, E-08028 Barcelona, Spain}
\author{M. Centelles}
\email{mariocentelles@ub.edu}
\affiliation{Departament de F\'isica Qu\`antica i Astrof\'isica and Institut de Ci\`encies del Cosmos (ICCUB), 
Facultat de F\'isica, Universitat de Barcelona, Mart\'i i Franqu\`es 1, E-08028 Barcelona, Spain}
\author{X. Vi\~nas}
\email{xavier@fqa.ub.edu}
\affiliation{Departament de F\'isica Qu\`antica i Astrof\'isica and Institut de Ci\`encies del Cosmos (ICCUB), 
Facultat de F\'isica, Universitat de Barcelona, Mart\'i i Franqu\`es 1, E-08028 Barcelona, Spain}
\author{T. R. Routray}
\email{trr1@rediffmail.com}
\affiliation{School of Physics, Sambalpur University, Jyotivihar-768 019, India}

\date{\today}

\begin{abstract}
The properties of the core-crust transition in neutron stars are investigated using effective nuclear 
forces of finite-range.
Special attention is paid to the
so-called dynamical method for locating the transition point, which, apart from 
the stability of the uniform nuclear matter against
clusterization, also considers contributions due to finite-size effects. In particular,
 contributions to the transition density and pressure from the direct and exchange energies are carefully
analyzed. To this end, finite-range forces of Gogny, Modified Gogny Interaction (MDI) and Simple Effective Interaction
(SEI) types are used in the numerical applications. The results from the dynamical
approach are compared with those from the popular thermodynamical method that
neglects the surface and Coulomb effects in the stability condition.
The dependence of the core-crust transition on the stiffness of the symmetry 
energy of the finite-range models 
is also addressed. Finally, we analyze the impact of the transition point on the 
mass, thickness and fraction of the moment of inertia of the neutron star crust. 
Prominent differences in these crustal properties of the star are found between 
using the transition point obtained with the dynamical method or the 
thermodynamical method. It is concluded that the core-crust transition needs to be 
ascertained as precisely as possible in order to have realistic estimates of the 
observed phenomena where the crust plays a significant role.
\end{abstract}

\maketitle
\section{Introduction}
Neutron stars (NSs) are the ultradense and compact remnants of massive stars after 
supernova explosions \cite{shapiro83,haensel07}, which
are composed of neutrons, protons, leptons and, eventually, other exotic particles.
NSs are also electrically neutral objects, and their different components are distributed in such
a way that they are locally in $\beta$-equilibrium. 
The structure of a NS consists of a
solid crust at low densities encompassing a homogeneous core in liquid phase. The density is maximum at the center, several times
the nuclear matter density ($\simeq$ 2.3 $\times$ 10$^{14}$ g cm$^{-3}$), and decreases with the distance
reaching a value of the terrestrial
iron ($\simeq$ 7.5 g cm$^{-3}$) at the surface of the star.
The external part of the star, i.e., its outer crust, consists of nuclei 
distributed in a solid body-centered-cubic (bcc) lattice permeated by a free electron gas. 
When the density increases, the nuclei in the crust become so neutron rich that neutrons start to drip from 
them. 
In this scenario, the lattice structure of nuclear clusters still remains, but now it is embedded in free 
electron and neutron gases. When the average density reaches a value about half of the nuclear 
matter saturation density, the lattice structure disappears due to energetic reasons and the system 
changes to a liquid phase. The boundary between the outer and inner crust is determined by nuclear masses, 
and corresponds to the neutron drip out density around $4\times 10^{11}$ g cm$^{-3}$ \cite{ruster06,hampel08}.
However, the transition density from the crust to
the core is much more uncertain and strongly model dependent 
\cite{baym70,kubis06,ducoin07,xu09a,xu09b,xu10a,xu10b,Moustakidis10,Moustakidis12,Seif14,routray16,gonzalez17}.

To determine the core-crust transition density from the crust side it is required to have a precise 
knowledge of the Equation of State (EOS) in this region of the star. This is a complicated 
task owing to the presence of neutron gas and the possible existence of complex structures in the
deep layers of the inner crust, where the nuclear clusters may adopt shapes different
from the spherical one (i.e., what is called ``pasta phases") in order to minimize the energy
\cite{baym71, lattimer95, shen98a, shen98b, douchin01, sharma15, Carreau19}.
Therefore, it is easier to investigate the core-crust transition from the core side. To this
end, one searches for the violation of the stability conditions of the homogeneous core under
small amplitude oscillations, which indicates the appearance of nuclear clusters and,  
consequently, the transition to the inner crust. There are different ways to determine the transition 
density from the core side, namely the thermodynamical method \cite{kubis06,xu09a,Moustakidis10,Moustakidis12,Seif14,routray16,gonzalez17}, the dynamical method
\cite{baym71,xu09a,xu09b,xu10a,xu10b,pethick95,ducoin07}, the random phase approximation 
\cite{horowitz01a, horowitz01b,carriere03} and the Vlasov equation method \cite{chomaz04,providencia06,
ducoin08a,ducoin08b,pais10}. In the low-density regime of the core near the crust the NS matter is composed of
neutrons, protons and electrons. 
In the thermodynamical approach, the stability is discussed in terms of the bulk properties of 
the EOS by imposing mechanical and chemical stability conditions, which set the boundaries of
 the core in the homogeneous case.  
On the other hand, in the dynamical method, one introduces density fluctuations
for neutrons, protons and electrons, which
can be expanded in plane waves of amplitude $A_i$ ($i=n, p, e$) and of wavevector ${\bf k}$.
These fluctuations may be induced by collisions, which transfer some momentum to the system.
Following Refs.~\cite{baym71,pethick95,ducoin08a,ducoin08b,xu09a} one writes the variation of 
the energy density generated by these fluctuations in terms of the energy curvature matrix 
$C^J$. 
The system is stable when the curvature matrix is convex, that is, the transition density is 
obtained from the condition $\vert C^J \vert$=0. 
This condition determines the dispersion relation $\omega(k)$ of the collective excitations 
of the system due to the perturbation and the 
instability will appear when this frequency $\omega(k)$ becomes imaginary. 
Thus, the dynamical method is more realistic than the thermodynamical approach, as it incorporates 
surface and Coulomb effects in the stability condition that are not taken into account in the 
thermodynamical method.
  
The dynamical method introduced in Ref.~\cite{baym71} assumes that the energy density functional
can be expressed as a sum of a homogeneous bulk part, a inhomogeneous term
depending on the gradients of the neutron and proton densities and the direct Coulomb energy. 
The zero-range Skyrme forces directly provide an energy density functional of this type. For these 
interactions, the core-crust transition density has been estimated using both the 
thermodynamical and dynamical methods \cite{xu09a}. In the same reference it is found that the 
transition density calculated with the thermodynamical approach is always higher than the value calculated with
with the dynamical method. On the other hand, an analysis \cite{xu09a} of the transition densities 
using 51 Skyrme forces shows a decreasing trend, roughly linear, when the 
slope of the symmetry energy at saturation (parameter $L$) increases, in agreement with
previous findings \cite{oyamatsu07}.

The estimate of the transition density using the thermodynamical method with finite-range
forces is rather straightforward, as only the energy density and its first and second
derivatives with respect to the neutron and proton densities are needed 
\cite{xu09a,gonzalez17,routray16}. 
However, to obtain the core-crust transition density using the dynamical method with
finite-range forces is more involved. This estimate has been discussed, to our 
knowledge, only in the particular case of the MDI interaction in Refs.~\cite{xu09a,xu10b}.

In the dynamical method using finite-range interactions one needs to extract the gradient 
corrections, which are encoded in the force but do not appear explicitly in the energy 
density functional in the case of finite-range interactions. 
To solve this problem, the authors in Ref.~\cite{xu09a} adopted the phenomenology of approximating the gradient contributions
with constant coefficients whose values are taken as the respective average values of the contributions
provided by 51-Skyrme interactions. 
A step further in the application of the dynamical method to estimate the core-crust 
transition density is discussed in Ref.~\cite{xu10b} also for the particular case of MDI
interactions. In this work the authors use the density matrix (DM) expansion proposed by 
Negele and Vautherin \cite{negele72a, negele72b} to derive a Skyrme-type energy density functional 
including gradient contributions, but with
density-dependent coefficients. Using this functional, the authors study the core-crust transition
through the stability conditions 
provided by the linealized Vlasov equations in NS matter.

The finite-range Gogny forces were proposed by D. Gogny almost forty years ago aimed to describe simultaneously
the mean-field and the pairing field \cite{decharge80, berger91, chappert08, goriely09}. These interactions are well suited for dealing with the deformation
properties of finite nuclei and to study fission barriers. However, Gogny interactions fail when one extrapolates 
to the neutron star domain \cite{Sellahewa14, gonzalez17} in spite that in the fitting protocol of the 
D1N~\cite{chappert08} and D1M~\cite{goriely09} interactions it is imposed that 
they reproduce the microscopic neutron matter EOS of Friedman and Pandharipande \cite{Friedman81}. To remedy this situation we have proposed \cite{gonzalez18} very recently
a reparametrization procedure of the D1M Gogny force, labeled D1M$^{*}$, able to reach a maximum neutron star mass of 2$M_\odot$ as 
required by recent astronomical observations~\cite{Demorest10, Antoniadis13}. 

The so-called Modified Gogny Interactions (MDI) \cite{das03}
were originally devised for transport calculations in heavy ion collisions,
but they have also been applied to other different scenarios (see \cite{xu09a} and references
therein), in particular to NS \cite{xu09a,xu09b,xu10a,xu10b}.
The MDI interactions can be rewritten as a zero-range contribution plus a finite range term
with a single Yukawa form factor. The parameters of the MDI interactions are 
determined subject to the constraint 
that the momentum-dependence of its single-particle potential reproduces, 
as best as possible, the predictions of the Gogny interaction \cite{das03}.

The SEI interaction is an effective nuclear force constructed with a minimum number of parameters to 
study the momentum and density dependence of the nuclear mean field \cite{behera98, Behera05}. This interaction has a 
single finite-range part, which can be of any conventional form, Gaussian, Yukawa or exponential, and 
two zero-range terms, one of them density-dependent, containing altogether eleven parameters. 
Unlike the case of effective
forces such as Skyrme, Gogny or M3Y, nine of the
eleven parameters of SEI are fitted using experimental or empirical constraints in nuclear
matter that allow one to reproduce the behaviour of the momentum and density dependence of the nuclear 
matter mean field as predicted by the microscopic Dirac-Brueckner-Hartree-Fock (DBHF), Brueckner-Hartree-Fock 
(BHF) and variational calculations (see
\cite{behera13} and references therein). One of the remaining two parameters is determined from the study of spin 
polarized matter \cite{behera15} and the other one along with the strength of the spin-orbit interaction, absent 
in nuclear matter calculations, is fitted to reproduce the experimental
binding energies of 620 even-even nuclei through HFB calculations \cite{behera98}.

Our aim in this work is to discuss again the core-crust transition in the case of finite-range interactions
enlarging the types of forces analyzed by including, in addition to MDI, the Gogny 
 and SEI interactions. On the other hand, in this work we will use a
DM expansion based on the Extended Thomas-Fermi approximation that was introduced in
Ref.~\cite{soubbotin00}. This DM expansion provides, using the Kohn-Sham method in the framework of the density
functional theory and including pairing correlations, a quasi-local energy density functional. With this functional 
one can compute in a simple way ground-state properties of finite nuclei in excellent agreement with the predictions 
of the full Hartree-Fock-Bogoliubov (HFB) calculations using finite-range interactions 
\cite{soubbotin03,krewald06,behera16}. The paper is organized as follows. In Section \ref{SecETF} the basic theory 
of the DM expansion used in this work is discussed. In Section \ref{SecTransition} we briefly review the most relevant
theoretical aspects of the dynamical method to compute the core-crust transition. Section \ref{results} is devoted 
to the discussion of the results obtained in our calculation and to the comparison with earlier estimates 
available in the literature. In Section \ref{summary} we give our summary and conclusions.
Technical details about the ETF approximation and the dynamical method for finite-range effective nuclear
forces are given in Appendices \ref{ApA} and \ref{AppB}, respectively.

\section{The Extended Thomas-Fermi approximation with finite-range forces}\label{SecETF}
The total energy density provided by a finite-range density-dependent effective nucleon-nucleon interaction can be decomposed as 
\begin{equation}
 \mathcal{H} = \mathcal{H}_{kin} +  \mathcal{H}_{zr} + \mathcal{H}_{dir} + \mathcal{H}_{exch}+ \mathcal{H}_{Coul}+\mathcal{H}_{LS},
 \label{eqendens}
\end{equation}
where $\mathcal{H}_{kin}$, $\mathcal{H}_{zr}$, $\mathcal{H}_{dir}$, $\mathcal{H}_{exch}$, $\mathcal{H}_{Coul}$ and $\mathcal{H}_{LS}$ are the
 kinetic, zero-range, finite-range direct, finite-range exchange, Coulomb and spin-orbit contributions.
The finite-range part of the interaction can be written in a general way as 
\begin{equation}\label{eqVfin}
V({\bf r},{\bf r'}) = \sum_m \left(W_m + B_m P^{\sigma} - 
H_m P^{\tau} -
M_m P^{\sigma} P^{\tau}\right) v_m({\bf r} , {\bf r'}),\\
\end{equation}
where $P^{\sigma}$ and $P^{\tau}$ are the spin and isospin exchange operators and $v_m({\bf r} , {\bf r'})$ are the form factors of the force. 
For Gaussian-type interactions the form factor is 
\begin{equation}
 v_m({\bf r} , {\bf r'})=e^{-|{\bf r} - {\bf r'}|^2/\alpha_m^2} ,
\end{equation}
 while for a Yukawa force one has\footnote{Note that the Yukawa form factor for MDI interactions reads as 
 $ v_m({\bf r} , {\bf r'})=\frac{e^{-\mu_m |{\bf r} - {\bf r'}|}}{|{\bf r} - {\bf r'}|}.$}
 
 \begin{equation}
  v_m({\bf r} , {\bf r'})=\frac{e^{-\mu_m |{\bf r} - {\bf r'}|}}{\mu_m |{\bf r} - {\bf r'}|}.
 \end{equation}
The finite-range term provides the direct ($\mathcal{H}_{dir}$) and exchange ($\mathcal{H}_{exch}$) 
contributions to the total energy density in Eq.~(\ref{eqendens}).

The HF energy due to the finite-range part of the interaction, neglecting zero-range, Coulomb and spin-orbit contributions, reads 
\begin{widetext}
\begin{equation}
E_{HF} = \sum_q \int d{\bf r} \left[\frac{\hbar^2}{2M}\tau({\bf r})
+ {\cal H}_{dir} + {\cal H}_{exch} \right]_q= \sum_q \int d{\bf r}\left[\frac{\hbar^2}{2M}\tau({\bf r}) + \frac{1}{2}\rho({\bf r})V^{H}({\bf r})
+ \frac{1}{2}\int d{\bf r'} V^{F}({\bf r},{\bf r'})\rho \left({\bf r} ,{\bf r'}\right)
\right]_q,
\label{eq2}
\end{equation}
\end{widetext}
where the subscript $q$ refers to each kind of nucleon. In this equation $\rho({\bf r})$ and $\tau({\bf r})$ 
are the particle and kinetic energy densities, respectively, and $\rho \left({\bf r} ,{\bf r'}\right)$ 
is the one-body density matrix.
The direct ($V^H$) and exchange ($V^F$) contributions to the HF potential due to the finite-range interaction (\ref{eqVfin}) are 
given by 
\begin{equation}\label{eq3}
V_q^H ({\bf r}) = \sum_m \int d{\bf r'} v_m ({\bf r} , {\bf r'}) \left[ D_{l, dir}^m \rho_q ({\bf r'}) + D_{u,dir}^m \rho_{q'} ({\bf r'})\right]
\end{equation}
and
\begin{eqnarray}
V_q^F({\bf r}, {\bf r'})&=& - \sum_m v_m ({\bf r} , {\bf r'}) \left[ D_{l, exch}^m \rho_q ({\bf r} ,{\bf r'})\right. \nonumber\\
&+&\left. D_{u,exch}^m \rho_{q'} ({\bf r},{\bf r'}) \right],
\label{eq4}
\end{eqnarray}
respectively. In Eqs.~(\ref{eq3}) and (\ref{eq4}) the coefficients $D_{l, dir}^m$, $D_{u,dir}^m$, $D_{l, exch}^m$ and $D_{u,exch}^m$ 
are the usual contributions of the spin
and isospin strengths for the like and unlike nucleons in the direct and exchange potentials:
\begin{eqnarray}
&&D_{L,dir}^m = W_m + \frac{B_m}{2} - H_m - \frac{M_m}{2} \nonumber \\
&&D_{U,dir}^m= W_m + \frac{B_m}{2}\nonumber\\
&&D_{L,exch}^m=M_m + \frac{H_m}{2} - B_m - \frac{W_m}{2}\nonumber\\
&&D_{U,exch}^m=M_m + \frac{H_m}{2}.
\end{eqnarray}

The non-local one-body DM $\rho({\bf r},{\bf r'})=\sum_i \varphi_i^*({\bf r})\varphi_i ({\bf r'})$
plays an essential role in Hartree-Fock (HF) calculations using effective finite-range forces,
where its full knowledge is needed. 
The HFB theory with finite-range forces is well
established from a theoretical point of view \cite{decharge80,nakada03} and calculations in
finite nuclei are feasible nowadays with a reasonable computing time. However, these calculations 
are still complicated and usually require specific codes, as for example the one provided by Ref.~\cite{robledo02}. 
Therefore, approximate methods based on the expansion of the DM in terms of local densities and their gradients usually allow one to
reduce the non-local energy density to a local form.

The simplest approximation to the DM is to replace locally its quantal value by its expression 
in nuclear matter, i.e., the so-called Slater or Thomas-Fermi (TF) approximation. A more elaborated
treatment was developed by Negele and Vautherin \cite{negele72a, negele72b}, which expanded the DM into a
bulk term (Slater) plus a corrective contribution that takes into account the finite-size effects.
Campi and Bouyssy \cite{campi78a,campi78b} proposed another approximation consisting of a Slater term alone but 
with an effective Fermi momentum, which partially resummates the finite-size corrective terms.
More recently, Soubbotin and Vi\~nas developed the extended Thomas-Fermi (ETF) approximation 
of the one-body DM in the case of a non-local single-particle Hamiltonian \cite{soubbotin00}. 
The ETF DM includes, on top of the Slater part, corrections of $\hbar^2$ order, which are
expressed through second-order derivatives of the neutron and proton densities. 
In the same Ref.~\cite{soubbotin00} the similarities and differences
with previous DM expansions \cite{negele72a, negele72b,campi78a,campi78b} are discussed in detail.

In this work we will use the ETF DM of Ref.~\cite{soubbotin00}. It can be written, for each type of nucleon, as 
\begin{equation}
 \rho\left({\bf R} + \frac{\bf s}{2},{\bf R} - \frac{\bf s}{2}\right) =\rho_0({\bf R}, {\bf s})+\rho_2({\bf R}, {\bf s}),
 \label{eq1}
\end{equation}
where ${\bf R}=({\bf r}+{\bf r'})/2$ and ${\bf s}={\bf r}-{\bf r'}$ are the center of mass and relative 
coordinates of the two nucleons, respectively. The TF contribution to the DM is given by
\begin{equation}\label{eqrho0}
 \rho_0\left({\bf R} ,{\bf s} \right)= \rho({\bf R}) \frac{3 j_1\big(k_F({\bf R})s\big)}{k_F({\bf R})s}
\end{equation}
where $j_1$ is the spherical Bessel function 
with $l=1$, $\rho({\bf R})$ is the local density and 
$k_F({\bf R})= \big(3\pi^2\rho({\bf R})\big)^{1/3}$ is the corresponding Fermi momentum. 
The contribution $\rho_2({\bf R} ,{\bf s})$ is the $\hbar^2$ contribution to
the DM, derived in Refs.~\cite{gridnev98,soubbotin00} and given explicitly in Eq.~(\ref{eqA1}) of Appendix \ref{ApA}.
The HF energy (\ref{eq2}) at ETF
level calculated using the DM given by Eq.~(\ref{eqA1}) reads
\begin{widetext}
\begin{equation}
E_{HF}^{ETF} = \sum_{q} \int d{\bf R}\left[\frac{\hbar^2}{2M}\frac{3}{5}(3\pi^2)^{2/3}\rho^{5/3} 
+ \frac{1}{2}\rho({\bf R})V^{H}({\bf R}) + \frac{1}{2} \int d{\bf s}  \rho_0 ({\bf R}, {\bf s}) V_0^F ({\bf R}, {\bf s}) 
+ \frac{\hbar^2 \tau_2({\bf R})}{2M} + {\cal H}_{exch,2}({\bf R})\right]_q,
\label{eq5}
\end{equation}
\end{widetext}
where $\tau_2({\bf R})$ and ${\cal H}_{exch,2}({\bf R})$
are the $\hbar^2$ contribution to the kinetic and exchange energies, which are explicitly given in 
Eqs.~(\ref{eqA41}) and (\ref{eqA6}), respectively,
 in Appendix~\ref{ApA}.
In Eq.~(\ref{eq5}) $V_0^F ({\bf R}, {\bf s})$ is the exchange potential for each kind of nucleon at TF level computed with
Eq.~(\ref{eq4}) using the Slater DM (\ref{eqrho0}), i.e.
\begin{widetext}
\begin{eqnarray}
V^{F}_{0}({\bf R},k) = - \sum_m \left[ D_{L,exch}^m \int d{\bf s}v(s)\frac{3j_1(k_{Fq} s)}{k_{Fq} s}\rho_q j_0(k s)\right.
+ \left.D_{U,exch}^m \int d{\bf s}v(s)\frac{3j_1(k_{Fq'} s)}{k_{Fq'} s}\rho_{q'} j_0(k s)\right].
\label{eqA9}
\end{eqnarray}
 \end{widetext}
From Eq.~(\ref{eq5}) we see that the energy in the ETF approximation for finite-range forces consists of a 
pure TF part, which depends only on the local densities of each type of particles, plus additional $\hbar^2$ 
corrections coming from the $\hbar$-expansion of the kinetic and exchange energy densities. 
As shown in Appendix \ref{ApA}, these $\hbar^2$ corrections for each type
of nucleon can be written, under an integral sign, as quadratic functions of the gradients of the neutron and 
proton densities with density-dependent coefficients:
\begin{eqnarray}
&&\int d{\bf R}\bigg[\frac{\hbar^2}{2M}\tau_2({\bf R}) + {\cal H}_{exch,2}({\bf R})\bigg]_q = \nonumber \\ 
&&\int d{\bf R}\bigg[B_{qq}(\rho_n,\rho_p)({\bf \nabla}\rho_q)^2
+ B_{qq'}(\rho_n,\rho_p){\bf \nabla}\rho_{q} \cdot {\bf \nabla}\rho_{q'}\bigg]_q, \nonumber\\
\label{eq5a}
\end{eqnarray}
where $q=n,p$ and $q'=p,n$, respectively (see Eqs.~(\ref{eqA10})--(\ref{eqA12}) of Appendix \ref{ApA}
for the explicit expressions of the $B_{qq}$ and $B_{qq'}$ coefficients).

\section{Dynamical Method: the neutron star core-crust transition}
\label{SecTransition}
We want to study the stability of NS matter against the formation of nuclear clusters. This matter is 
composed of neutrons, protons and electrons. It is globally charge neutral and satisfies the $\beta$-equilibrium 
condition \cite{shapiro83, haensel07}.
To investigate the core-crust transition, we generalize 
the formalism developed by Baym, Bethe and Pethick \cite{baym71} to the case of finite-range forces. 
Following this reference, we write the particle density as a unperturbed constant part plus a 
position-dependent fluctuating contribution:
\begin{equation}
\rho({\bf R})= \rho_0 + \delta \rho ({\bf R}).
\label{eq6}
\end{equation}
Using (\ref{eq6}), the total energy of the system can be expanded as:
\begin{widetext}
\begin{eqnarray}
E &=& E_0 + \int d{\bf R} \left[\frac{\partial {\cal H}}{\partial \rho_{n}}\delta \rho_n +
\frac{\partial {\cal H}}{\partial \rho_{p}}\delta \rho_p + \frac{\partial {\cal H}}{\partial \rho_{e}}\delta \rho_e \right]_{\rho_0}
+ \frac{1}{2}\int d{\bf R} \left[\frac{\partial^2 {\cal H}}{\partial \rho_{n}^2}(\delta \rho_n)^2 
+ \frac{\partial^2 {\cal H}}{\partial \rho_{p}^2}(\delta \rho_p)^2 + \frac{\partial^2 {\cal H}}{\partial \rho_{e}^2}
(\delta \rho_e)^2 + 2\frac{\partial^2 {\cal H}}{\partial \rho_{n} \partial \rho_{p}}\delta \rho_n \delta \rho_p \right]_{\rho_0}
\nonumber \\ 
&+& \int d{\bf R}\left[B_{nn}(\rho_{n},\rho_{p})\left({\bf \nabla}\delta \rho_n\right)^2 
+ B_{pp}(\rho_{n},\rho_{p})\left({\bf \nabla}\delta \rho_p\right)^2 
+ B_{np}(\rho_{n},\rho_{p}){\bf \nabla}\delta \rho_n\cdot{\bf \nabla}\delta \rho_p 
+ B_{pn}(\rho_{p},\rho_{n}){\bf \nabla}\delta \rho_p\cdot{\bf \nabla}\delta \rho_n \right]_{\rho_0}
\nonumber \\
&+& \int d{\bf R}\left[{\cal H}_{dir}(\delta \rho_n,\delta \rho_p) + {\cal H}_{Coul}(\delta \rho_p,\delta \rho_e)\right],
\label{eq7}
\end{eqnarray}
\end{widetext}
where $E_0$ contains the contribution to the energy from the unperturbed parts of the neutron, 
proton and electron densities, $\rho_{0n}$, $\rho_{0p}$ and $\rho_{0e}$, respectively. The subscript 
$\rho_0$ labeling the square brackets in Eq.~(\ref{eq7}) implies that the derivatives of the energy 
density ${\cal H}$ as well as the coefficients $B_{qq}$ are 
evaluated at the unperturbed nucleon and electron densities. The last integral in Eq.~(\ref{eq7}) is the 
contribution from the nuclear direct and Coulomb parts arising out of the fluctuation in 
the particle densities. The two terms of this last integral in Eq.~(\ref{eq7}) are explicitly given by
\begin{widetext}
\begin{equation}\label{Hdir}
 {\cal{H}}_{dir}  (\delta \rho_n, \delta \rho_p) = \frac{1}{2} \sum_q \delta \rho_q ({\bf R}) \int d {\bf s} \left[  \sum_m D_{L,dir}^{m} v_m ({\bf s}) 
 \delta \rho_q ({\bf R} - {\bf s}) + \sum_m D_{U,dir}^m v_m ({\bf s}) \delta \rho_{q'} ({\bf R} - {\bf s}) \right]
\end{equation}
and 
 \begin{equation}
  {\cal{H}}_{Coul}  (\delta \rho_n, \delta \rho_p) = \frac{e^2}{2}  (\delta \rho_p  ({\bf R})- \delta \rho_e  ({\bf R})) \int d {\bf s} 
  \frac{\delta \rho_p ({\bf R} - {\bf s} ) -\delta \rho_e ({\bf R} - {\bf s} ) }{s}.
 \end{equation}
\end{widetext} 
Linear terms in the $\delta\rho$ fluctuation vanish in Eq.~(\ref{eq7}) by the following reason. We are assuming that neutrons, protons and
electrons are in $\beta$-equilibrium. Therefore, the corresponding chemical potentials, defined as
$\mu_i={\partial {\cal H}}/{\partial \rho_{i}}\vert_{\rho_0}$ for each kind of particle ($i=n,p,e$), 
fulfill $\mu_n - \mu_p = \mu_e$. Using this fact, the linear terms in Eq.~(\ref{eq7}) can be written as:
\begin{equation}
\mu_n \delta \rho_n + \mu_p \delta \rho_p + \mu_e \delta \rho_e =
\mu_n( \delta \rho_n + \delta \rho_p ) + \mu_e( \delta \rho_e - \delta \rho_p ).
\label{eq7a}
\end{equation}
The integration of this expression over the space vanishes owing to the charge neutrality of the matter 
(i.e., $\int d{\bf R} (\delta \rho_e - \delta \rho_p)=0$) and to the conservation of the baryon number 
(i.e., $\int d{\bf R} (\delta \rho_n + \delta \rho_p)=0$).

Next, we write the varying particle densities as the Fourier transform of 
the corresponding momentum distributions $\delta n_q({\bf k})$ as \cite{baym71}
\begin{equation}
\delta \rho_q({\bf R}) = \int \frac{d{\bf k}}{(2\pi)^3} \delta n_q({\bf k}) e^{i {\bf k} \cdot {\bf R}}.
\label{eq8}
\end{equation}
One can transform this equation to momentum space due 
to the fact that the fluctuating densities are the only quantities in Eq.~(\ref{eq7}) that depend on the position.
Consider for example the crossed gradient term ${\bf \nabla}\delta \rho_n\cdot{\bf \nabla}\delta \rho_p$
in Eq.~(\ref{eq7}).
Taking into account Eq.~(\ref{eq8}) we can write 
\begin{widetext}
\begin{eqnarray}
\int d{\bf R}{\bf \nabla}\delta \rho_n\cdot{\bf \nabla}\delta \rho_p &=&
-\int \frac{d{\bf k_1}}{(2 \pi)^3}\frac{d{\bf k_2}}{(2 \pi)^3}{\bf k_1}\cdot{\bf k_2} \delta n_n({\bf k_1}) \delta n_p({\bf k_2})
\int d{\bf R}e^{i({\bf k_1}+{\bf k_2})\cdot{\bf R}}\nonumber\\
&=&-\int \frac{d{\bf K}d{\bf k}}{(2 \pi)^3} \left(\frac{{\bf K}}{2}+{\bf k}\right) \cdot \left(\frac{{\bf K}}{2}-{\bf k}\right)
\delta n_n\left(\frac{{\bf K}}{2}+{\bf k}\right) \delta n_p\left(\frac{{\bf K}}{2}-{\bf k}\right)
\delta({\bf K})\nonumber \\
&=& \int \frac{d{\bf k}}{(2 \pi)^3}\delta n_n({\bf k}) \delta n_p(-{\bf k})k^2 =
\int \frac{d{\bf k}}{(2 \pi)^3} \delta n_n({\bf k})\delta n_p^*({\bf k})k^2, 
\label{eqB1}
\end{eqnarray}
\end{widetext}
where we have used the fact that $\delta \rho_q = \delta\rho^* _q$ and, therefore, due to (\ref{eq8}), 
$\delta n_q(-{\bf k}) = \delta n^*_q({\bf k})$. Similarly,
the other quadratic terms in the fluctuating density in Eq.~(\ref{eq7}) can also be transformed 
into integrals in momentum space of quadratic combinations of fluctuations of the momentum distributions (\ref{eq10}).
After some algebra, one obtains
\begin{widetext}
\begin{eqnarray}
E &=& E_0 +  \frac{1}{2}\int \frac{d{\bf k}}{(2\pi)^3} 
\left\{\left[\frac{\partial \mu_n}{\partial \rho_n}\delta n_n({\bf k})\delta n^*_n({\bf k})
+ \frac{\partial \mu_p}{\partial \rho_p}\delta n_p({\bf k})\delta n^*_p({\bf k})
+ \frac{\partial \mu_n}{\partial \rho_p}\delta n_n({\bf k})\delta n^*_p({\bf k})
+\frac{\partial \mu_p}{\partial \rho_n}\delta n_p({\bf k})\delta n^*_n({\bf k})\right.\right.\nonumber\\
&+& \left.\left.
 \frac{\partial \mu_e}{\partial \rho_e}\delta n_e({\bf k})\delta n^*_e({\bf k})\right]_{\rho_0}\right.
\nonumber \\
&+& 2k^2\left[B_{nn}(\rho_{n},\rho_{p})\delta n_n({\bf k})\delta n^*_n({\bf k})
+ B_{pp}(\rho_{n},\rho_{p})\delta n_p({\bf k})\delta n^*_p({\bf k})
+ B_{np}(\rho_{n},\rho_{p})\left(\delta n_n({\bf k})\delta n^*_p({\bf k}) 
+ \delta n_p({\bf k})\delta n^*_n({\bf k})\right)\right]_{\rho_0}
\nonumber \\
&+& \sum_m\left[D_{L,dir}^m\left(\delta n_n({\bf k})\delta n^*_n({\bf k}) + \delta n_p({\bf k})\delta n^*_p({\bf k})\right)
+ D_{U,dir}^m\left(\delta n_n({\bf k})\delta n^*_p({\bf k})+\delta n_p({\bf k})\delta n^*_n({\bf k})\right)
\right]({\cal F}_m(k)- {\cal F}_m(0)) 
\nonumber \\
&+&\left. \frac{4 \pi e^2}{k^2}\left(\delta n_p({\bf k})\delta n^*_p({\bf k}) + \delta n_e({\bf k})\delta n^*_e({\bf k})
- \delta n_p({\bf k})\delta n^*_e({\bf k}) - \delta n_e({\bf k})\delta n^*_p({\bf k})\right)\right\}.
\label{eq9}
\end{eqnarray}
\end{widetext}
The factors ${\cal F}_m(k)$ which enter in the contributions of the direct potential 
in Eq.~(\ref{eq9}) are the Fourier transform of the form factors
 $v_m(s)$ (see Appendix \ref{AppB} for more details).
We see that Eq.~(\ref{eq9}) can be expressed in a compact form as
\begin{equation}
E = E_0 + \frac{1}{2} \sum_{i,j} \int \frac{d{\bf k}}{(2\pi)^3}
\frac{\delta^2 E}{\delta n_{i}({\bf k}) \delta n_{j}^*({\bf k})} \delta n_{i}({\bf k}) \delta n_{j'}^*({\bf k}),
\label{eq10}
\end{equation}
where $E_0$ is the unperturbed energy and the subscripts 
$i$ and $j$ concern the different type of particle. 

The curvature matrix $C^f$ is given by
\begin{equation}
C^f =\frac{\delta^2 E}{\delta n_{i}({\bf k}) \delta n_{j}^*({\bf k})},
\label{eq11}
\end{equation}
which can be written as a sum of three matrices as
\begin{widetext}
\begin{eqnarray}
C^f = \left( \begin{array}{ccc}
\partial \mu_n / \partial \rho_n & \partial \mu_n / \partial \rho_p & 0 \\
 \partial \mu_p / \partial \rho_n& \partial \mu_p / \partial \rho_p & 0 \\
0 & 0 & \partial \mu_e / \partial \rho_e \end{array} \right)
 +  \left( \begin{array}{ccc}
D_{nn}(\rho, k) & D_{np}(\rho, k) & 0 \\
 D_{pn}(\rho,k) & D_{pp}(\rho, k) & 0 \\
0 &  0& 0 \end{array} \right) + \frac{4 \pi e^2}{k^2} \left( \begin{array}{ccc}
0 & 0 &  0 \\
0 & 1 & -1 \\
0 &-1 &  1 \end{array} \right), \nonumber
\\
\label{eq12}
\end{eqnarray}
\end{widetext}
where it is understood that all quantities are evaluated at the unperturbed densities.
The curvature matrix $C^f$ is composed of three different pieces. The first one, which is the dominant term, corresponds
to the bulk contribution. It defines the stability of uniform NS matter, and corresponds to
the equilibrium condition of the thermodynamical method for locating the core-crust transition point. The second piece describes the 
contributions due to the gradient expansion of the energy density functional. Finally, the last piece
 is due to the direct Coulomb interactions of protons and electrons. These last two terms
 tend to stabilize the system reducing the instability region predicted by the bulk contribution alone.
The functions $D_{qq'}(\rho,k)$ contain the terms of the nuclear energy density coming from the form factor
of the nuclear interaction plus the $\hbar^2$ contributions of the kinetic energy and exchange energy densities. They
can be written as:
\begin{small}
\begin{eqnarray}
 D_{nn}(\rho, k)&=& \sum_m D_{L,dir}^m\big({\cal F}_m(k)-{\cal F}_m(0)\big) + 2k^2 B_{nn}(\rho_{n},\rho_{p})\nonumber \\
 D_{pp}(\rho, k)&=& \sum_m D_{L,dir}^m\big({\cal F}_m(k)-{\cal F}_m(0)\big) + 2k^2 B_{pp}(\rho_{n},\rho_{p})\nonumber \\
 D_{np}(\rho, k)&=&  D_{pn}(\rho, k) \nonumber \\
&=&\sum_m D_{U,dir}^m\big({\cal F}_m(k)-{\cal F}_m(0)\big) + 2k^2 B_{np}(\rho_{n},\rho_{p}).\nonumber\\
\label{eq13}
\end{eqnarray}
\end{small}
The stability of the system against small density fluctuations requires that the curvature matrix $C^f$ has to be convex for all values
of $k$, and if this condition is violated, the system becomes unstable. 
The convexity of $C^f$ is guaranteed if the 3$\times$3 determinant of the matrix is positive, provided that 
$\partial \mu_n/\partial \rho_n$ (or $\partial \mu_p/\partial \rho_p$) and the 2$\times$2 minor of the nuclear sector in (\ref{eq12}) 
are also positive \cite{xu09a}. Therefore, the stability condition against cluster formation, which indicates the transition from
the core to the crust, is given by the condition that the dynamical potential, defined as 
\begin{widetext}
\begin{equation}
 V_{\mathrm{dyn}} (\rho, k) = \left(\frac{\partial \mu_p}{\partial \rho_p} + D_{pp}(\rho, k) + \frac{4 \pi e^2}{k^2}\right) 
 - \frac{\left( \partial \mu_n / \partial \rho_p + D_{np}(\rho, k)\right)^2}{\partial \mu_n / \partial \rho_n + D_{nn}(\rho, k)} 
- \frac{(4\pi e^2 / k^2)^2}{\partial \mu_e / \partial \rho_e + 4 \pi e^2 / k^2},
\label{eq14}
\end{equation}
\end{widetext}
be positive. For a given density, the dynamical potential $V_\mathrm{dyn}(\rho, k)$ is calculated at the $k$ value that 
minimizes Eq.~(\ref{eq14}), i.e., that fulfills
$\left.\partial V_\mathrm{dyn}(\rho, k) / \partial k \right|_\rho = 0$ \cite{baym71,ducoin07,xu09a}.
The first term of (\ref{eq14}) gives the stability of protons, and the second and third ones give the
stability when the protons interact with neutrons and electrons, respectively.
From the condition $\left.\partial V_\mathrm{dyn}(\rho, k) / \partial k \right|_\rho $ 
one gets the dependence of the momentum on the density, $k (\rho)$,
and therefore Eq.~(\ref{eq14}) becomes a function only depending on the unperturbed density $\rho=\rho_0$. 
The core-crust transition takes place at the density where the dynamical 
potential vanishes, i.e., $V_\mathrm{dyn} (\rho, k(\rho))=0$.

We conclude this section by mentioning that dropping the Coulomb contributions in Eq.~(\ref{eq14})
and taking the $k\to0$ limit leads to the so-called thermodynamical potential used in the thermodynamical method
for looking for the core-crust transition:
\begin{equation}
V_{\mathrm{ther}}(\rho) = \frac{\partial \mu_p}{\partial \rho_p} 
 - \frac{\left( \partial \mu_n / \partial \rho_p \right)^2}{\partial \mu_n / \partial \rho_n } ,
\label{eqB7}
\end{equation}
where the condition of stability of the uniform matter of the core corresponds to $V_{\mathrm{ther}}(\rho) > 0$.

\section{Numerical results and discussions}\label{results}
We will discuss first the impact of the finite-range terms of the nuclear effective force
on the dynamical potential. Next, we will focus on the study of the core-crust transition density 
and pressure calculated from the dynamical and thermodynamical methods using three different finite-range nuclear 
models. Finally, we will examine the influence of the core-crust transition point on the crustal properties in neutron stars.
\subsection{The dynamical potential}
\begin{figure}[t]
\centering
\includegraphics[clip=true,width=\columnwidth]{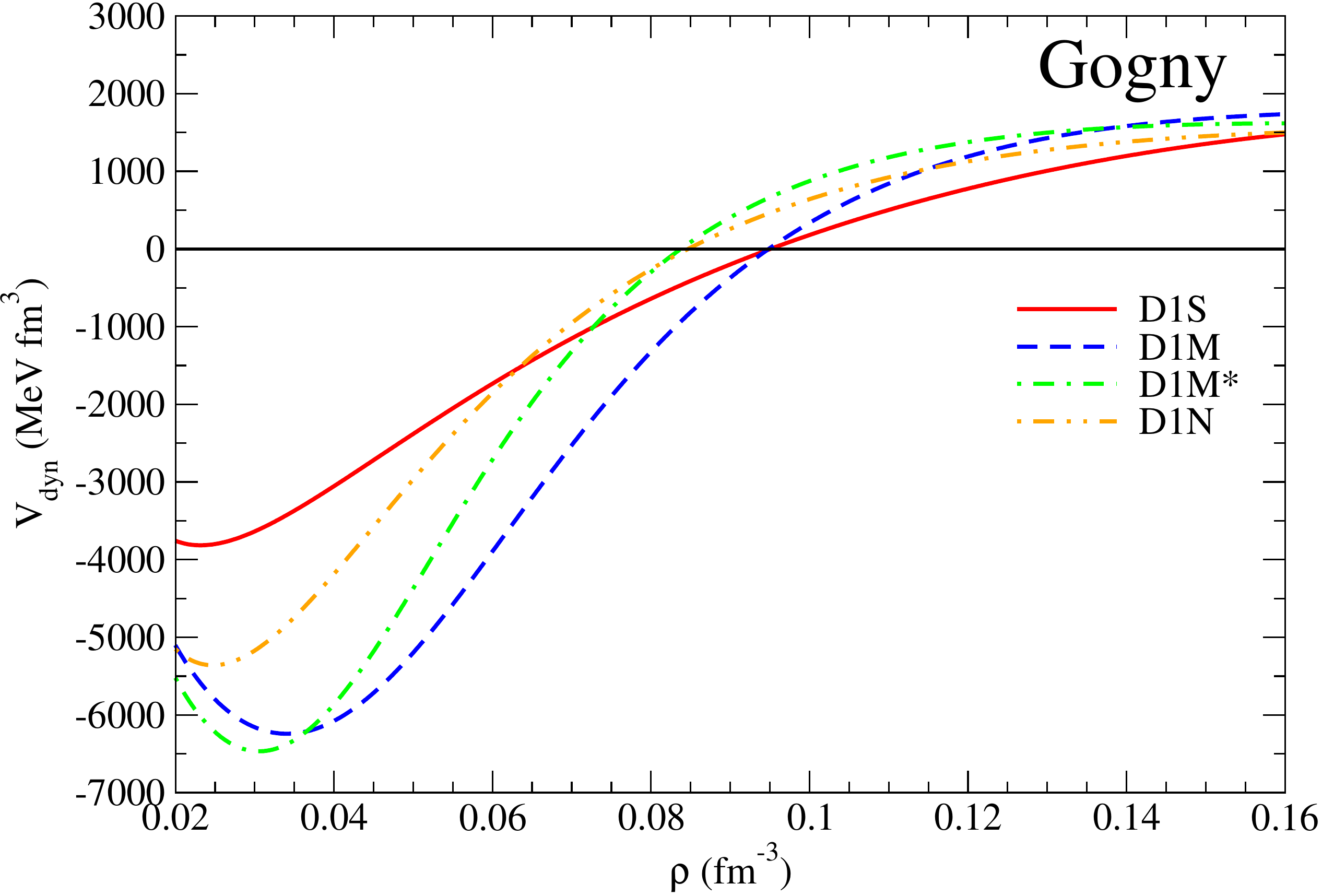}
\caption{Dynamical potential as a function of the density for the D1S, D1M, D1M$^*$ and D1N Gogny interactions. }\label{fig:Vdyn}
\end{figure}

We show in Fig.~\ref{fig:Vdyn} the dynamical potential as a function of the unperturbed density computed using the Gogny
forces D1S \cite{berger91}, D1M \cite{goriely09}, D1M$^*$ \cite{gonzalez18} and D1N \cite{chappert08}.
The density at which $V_\mathrm{dyn}(\rho, k(\rho))$ vanishes depicts the transition density. 
Below this point a negative value of $V_\mathrm{dyn}$  implies instability. Above this density, 
the dynamical potential is positive, meaning that the curvature matrix in Eq.~(\ref{eq12}) is convex at all 
values of $k$ and therefore the system is stable against cluster formation.
The Gogny forces D1S and D1M, predict larger values of the transition density compared to D1N and D1M$^*$. 
This is due to the different density slope of the symmetry energy predicted 
by these forces.
The values of the slope parameter $L$ of the symmetry energy 
\begin{equation}
 L =\left. 3 \rho_\mathrm{sat} \frac{\partial E_\mathrm{sym} (\rho)}{\partial \rho}\right|_{\rho_\mathrm{sat}}
\end{equation}
are $22.43$ MeV, $24.83$ MeV, $35.58$ MeV and $43.18$ MeV for the D1S, D1M, D1N and D1M* forces. 
A lower $L$ value implies a softer symmetry energy around the saturation density.
The prediction of higher core-crust
transition density for lower $L$ found in this study is in agreement with the previous 
results reported in the literature (see \cite{xu09a,gonzalez17} and references therein).

As has been done in previous works \cite{baym70,baym71,xu09a}, the dynamical potential (\ref{eq12}) can be
approximated up to order $k^2$ by performing a Taylor expansion of the coefficients $D_{nn}$, $D_{pp}$ and $D_{np}=D_{pn}$ in Eqs.~(\ref{eq13}). 
In this case the dynamical potential can be formally written as \cite{pethick95,ducoin07,xu09a}
\begin{equation}
{\tilde V}_{\mathrm{dyn}}(\rho, k) = V_{\mathrm{ther}}(\rho) + \beta(\rho)k^2 + 
\frac{4\pi e^2}{k^2 + \frac{4\pi e^2}{\partial \mu_e/ \partial \rho_e}},
\label{eq14a}
\end{equation}
where $V_{\mathrm{ther}}(\rho)$ has been defined in Eq.~(\ref{eqB7})
and the expression of $\beta (\rho)$ is given in Eq.~(\ref{eqB8}) of 
Appendix \ref{AppB}. 
The practical advantage of Eq.~(\ref{eq14a}) is that the $k$-dependence is separated from the $\rho$-dependence,  
whereas in the full expression for the dynamical potential in Eq.~(\ref{eq14}) they are not separated.
It may be mentioned that in the case of Skyrme forces, which are zero-range interactions,
the functions $D_{qq'}(\rho,k)$ of Eqs.~(\ref{eq13}) and (\ref{eq14})
are quadratic functions of the momentum $k$ with constant coefficients.
\begin{figure}[b]
\centering
\includegraphics[clip=true, width=\columnwidth]{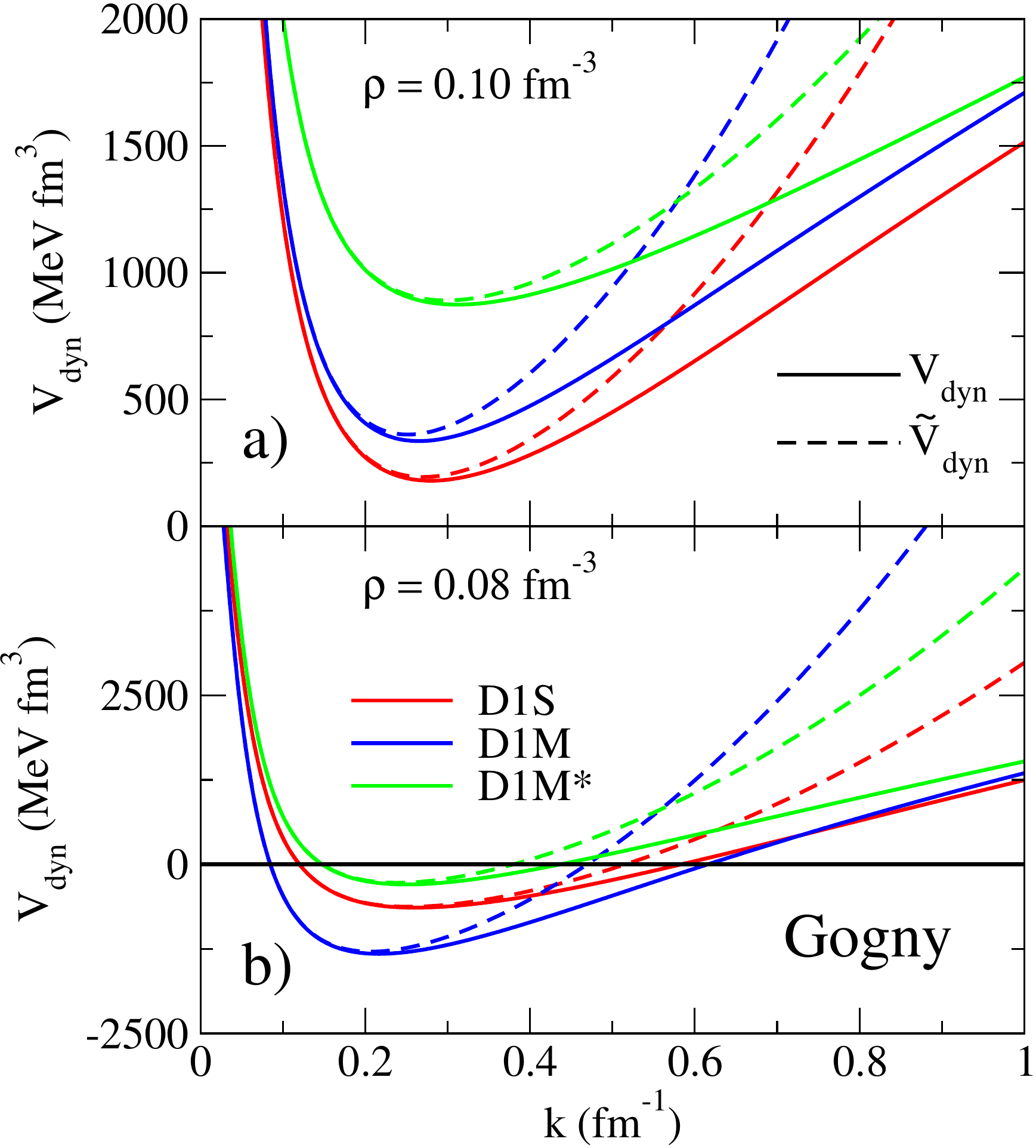}
\caption{Momentum dependence of the dynamical potential at two different densities, $\rho=0.10$ fm$^{-3}$ (panel a)
and $\rho=0.08$ fm$^{-3}$ (panel b), for the D1S, D1M and D1M$^{*}$ Gogny interactions. The results plotted with solid 
lines are obtained using the full expression of the dynamical potential, given in Eq.~(\ref{eq14}), while the dashed lines are 
the results of its $k^2$-approximation, given in Eq.~(\ref{eq14a}).}\label{fig:Vdyncomp}
\end{figure}

In order to examine the validity of the $k^2$-approximation (long-wavelength limit) of the dynamical potential,
we plot in Fig.~\ref{fig:Vdyncomp} the dynamical potential at a given density as a function of the momentum $k$ 
for both Eq.~(\ref{eq14}) (solid lines)
and its $k^2$-approximation in Eq.~(\ref{eq14a}) (dashed lines) for the Gogny forces D1S, 
D1M and D1M$^*$. 
The momentum dependence 
of the dynamical potential at density $\rho=0.10$ fm$^{-3}$ is shown in the upper panel of Fig.~\ref{fig:Vdyncomp}. 
One sees that at this density the core
has not reached the transition point for any of the considered Gogny forces, as the minima of the 
$ V_{\mathrm{dyn}} (\rho, k)$ curves are positive. This implies that the system is stable against formation of clusters.
In the lower panel of Fig.~\ref{fig:Vdyncomp} the same results but at a density $\rho=0.08$ fm$^{-3}$ are shown.
As the minimum of the dynamical potential for all forces is negative, the matter is unstable against 
cluster formation. From the results in Fig.~\ref{fig:Vdyncomp} it can be seen
that at low values of $k$ the agreement between the results of Eq.~(\ref{eq14}) and its approximation Eq.~(\ref{eq14a}) 
is very good. However, at large momenta beyond $k$ of the minimum of the dynamical potential, 
there are increasingly larger differences between both calculations of the dynamical potential. 
This is consistent with the fact that Eq.~(\ref{eq14a}) is the $k^2$-approximation of Eq.~(\ref{eq14}).

\begin{figure}[t]
\centering
\includegraphics[clip=true, width=\columnwidth]{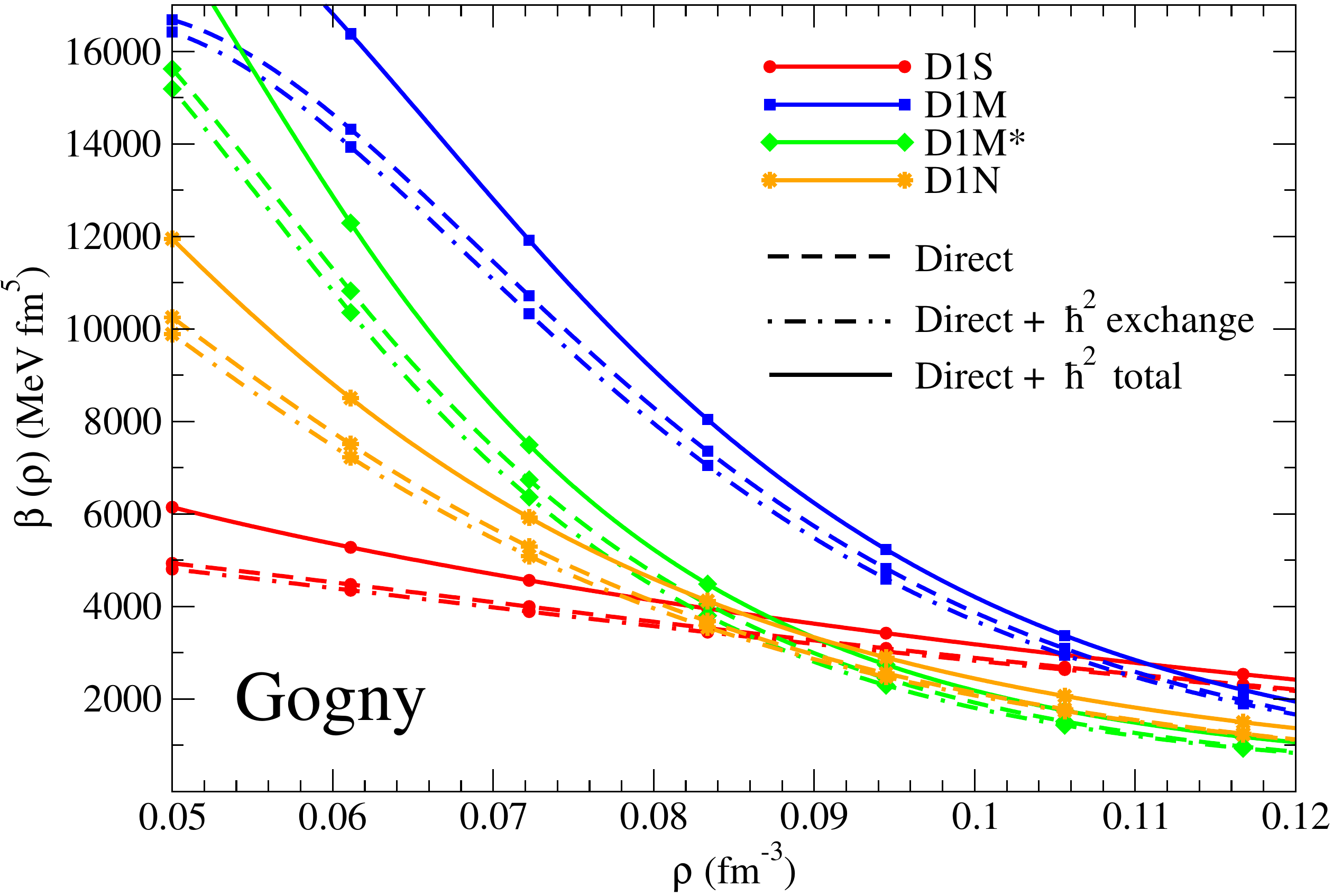}
\caption{Coefficient $\beta(\rho)$ of Eq.~(\ref{eq14a}) as a function of the density for a set of Gogny interactions. The dashed line  
includes only contributions coming from the direct energy, the dash-dotted lines include contributions coming 
from the direct energy and from the from the gradient effects of the exchange energy. 
Finally, the solid lines include all contributions to $\beta(\rho)$.}\label{fig:beta}
\end{figure}
 
To investigate the impact of the direct and $\hbar^2$ contributions of the finite-range part of the interaction
on the dynamical potential, in Fig.~\ref{fig:beta} we plot for Gogny forces
the behaviour of the coefficient $\beta(\rho)$, which accounts for the finite-size effects
in the stability condition of the core using the dynamical potential ${\tilde V}_{\mathrm{dyn}}(\rho, k)$.
The dashed lines are the result for $\beta (\rho)$ when only the direct contribution from the
finite range is taken into account.
The dash-dotted lines give the value of $\beta(\rho)$ where, along with the direct 
contribution, the gradient effects from the 
exchange energy are included. Finally, solid lines correspond to 
the total value of $\beta(\rho)$, which includes the direct contribution and
the complete $\hbar^2$ corrections, coming from both the exchange and kinetic energies.
 We see that for Gogny forces the gradient correction due to the 
$\hbar^2$ expansion of the exchange energy reduces the result of $\beta (\rho)$ from the direct contribution, at most, by 5\%. When the  
gradient corrections from both the exchange and kinetic energies are taken together, 
the total result for $\beta (\rho)$ increases by 10\% at most with respect to the direct contribution. The effects of the 
kinetic energy $\hbar^2$ corrections are around three times larger than the effects from the exchange energy $\hbar^2$ corrections 
and go in the opposite sense. 
Therefore, we conclude that the coefficient $\beta(\rho)$ is governed dominantly by the gradient expansion of the
direct energy (see Appendix \ref{AppB}), whereas the contributions from the $\hbar^2$ corrections to the 
exchange and kinetic energies are small corrections.

\subsection{The core-crust transition with finite-range effective interactions} 
\begin{figure}[b]
\centering
\includegraphics[clip=true,width=\columnwidth]{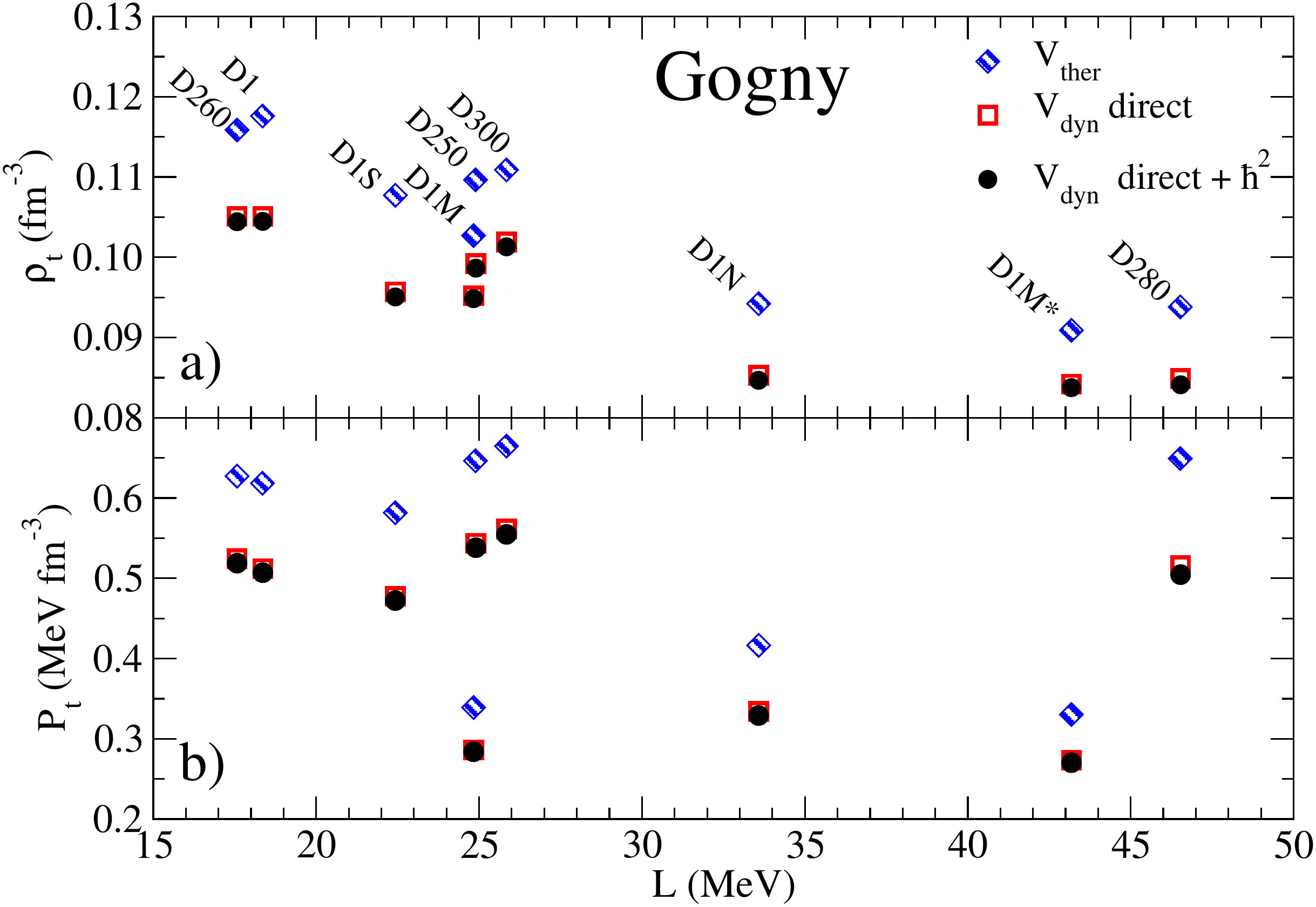}
\caption{Transition density (panel a) and transition pressure (panel b) as a function of the slope of the symmetry 
energy calculated using the thermodynamical and the dynamical methods for 
a set of different Gogny interactions.}\label{fig:rhotPtGogny}
\end{figure}

\begin{figure}[t]
\centering
\includegraphics[clip=true,width=\columnwidth]{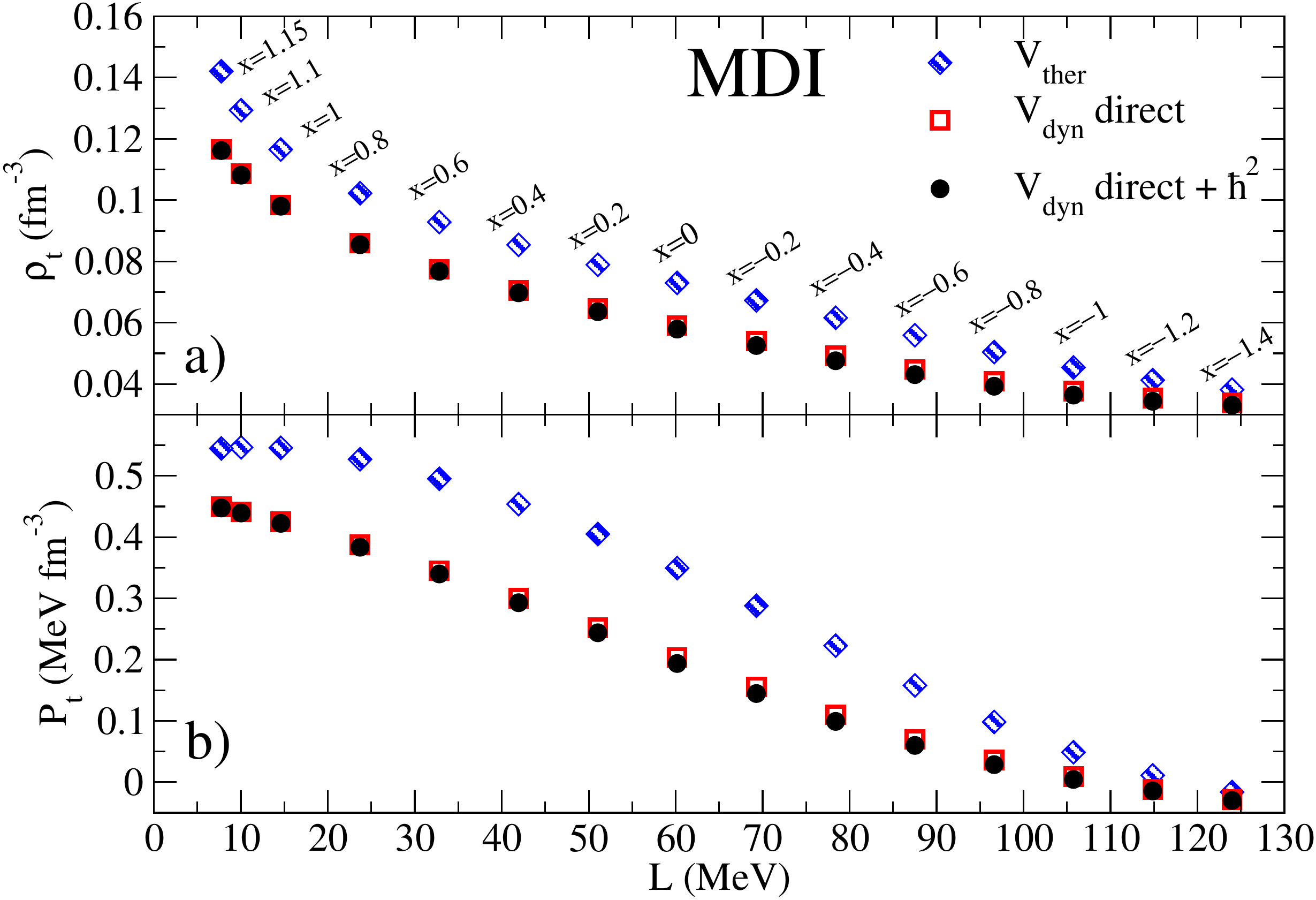}
\caption{Same as Fig.~\ref{fig:rhotPtGogny} but for a family of MDI interactions.}\label{fig:rhotPtMDI}
\end{figure}

\begin{figure}[t]
\centering
\includegraphics[clip=true, width=\columnwidth]{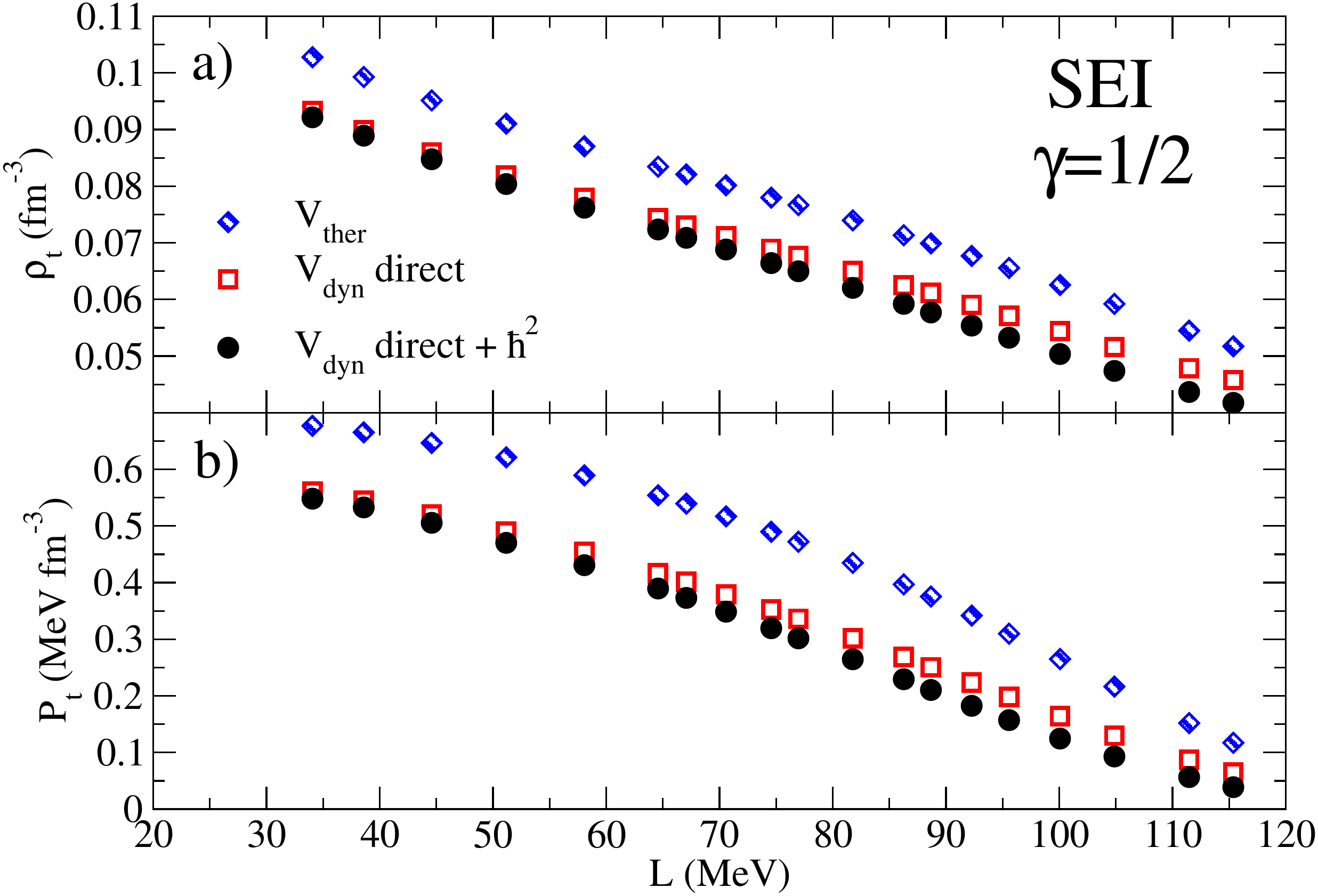}
\caption{Same as Fig.~\ref{fig:rhotPtGogny} but for a family of SEI interactions of $\gamma=1/2$ ($K_0=237.5$ MeV).}\label{fig:rhotPtSEI}
\end{figure}

\begin{figure}[h]
\centering
\includegraphics[clip=true, width=\columnwidth]{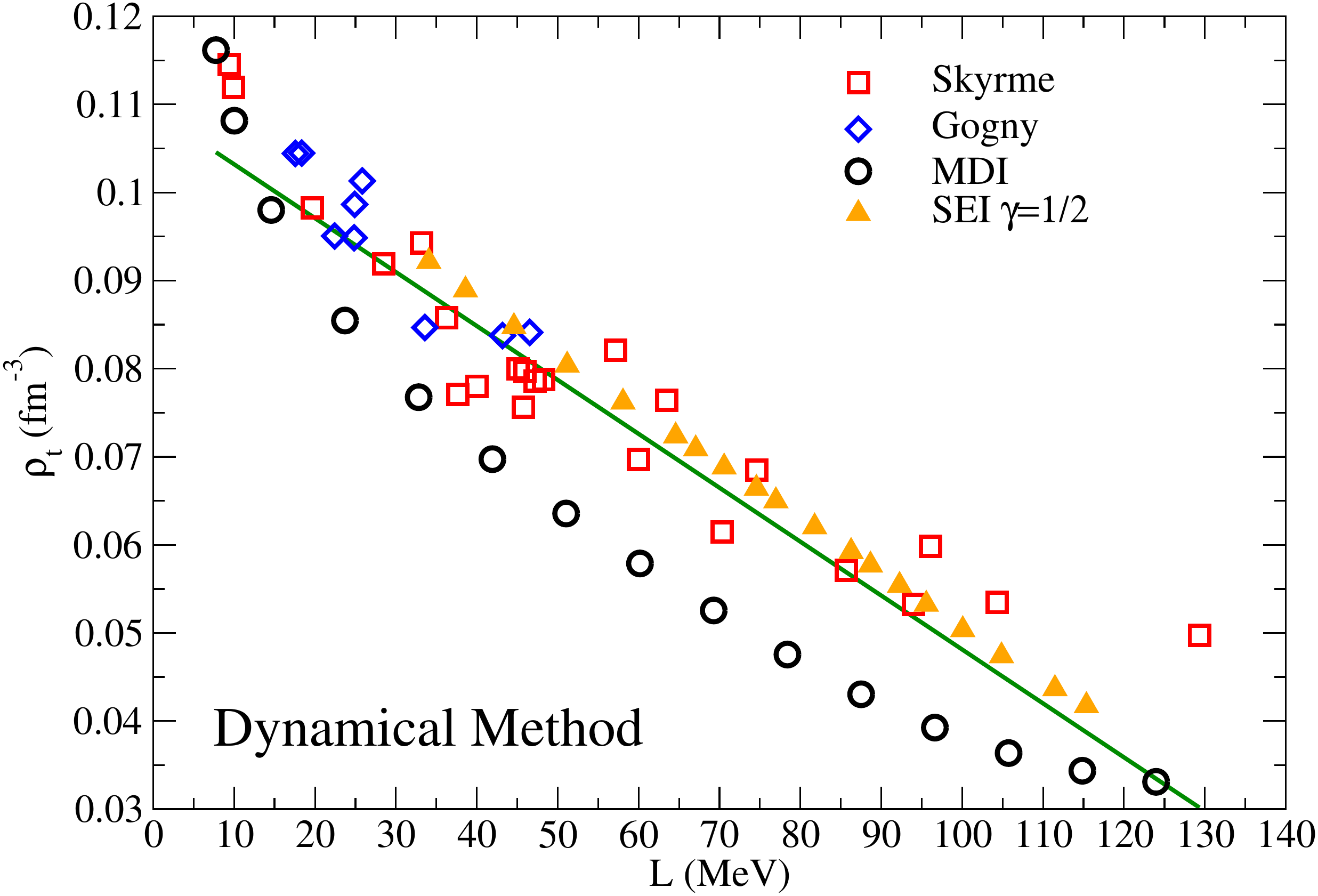}
\caption{Transition density as a function of the slope of the symmetry energy calculated using the dynamical method for 
a set of Skyrme, Gogny, MDI and SEI interactions.}\label{fig:rhotall}
\end{figure}
\begin{figure}[h]
\centering
\includegraphics[clip=true,width=\columnwidth]{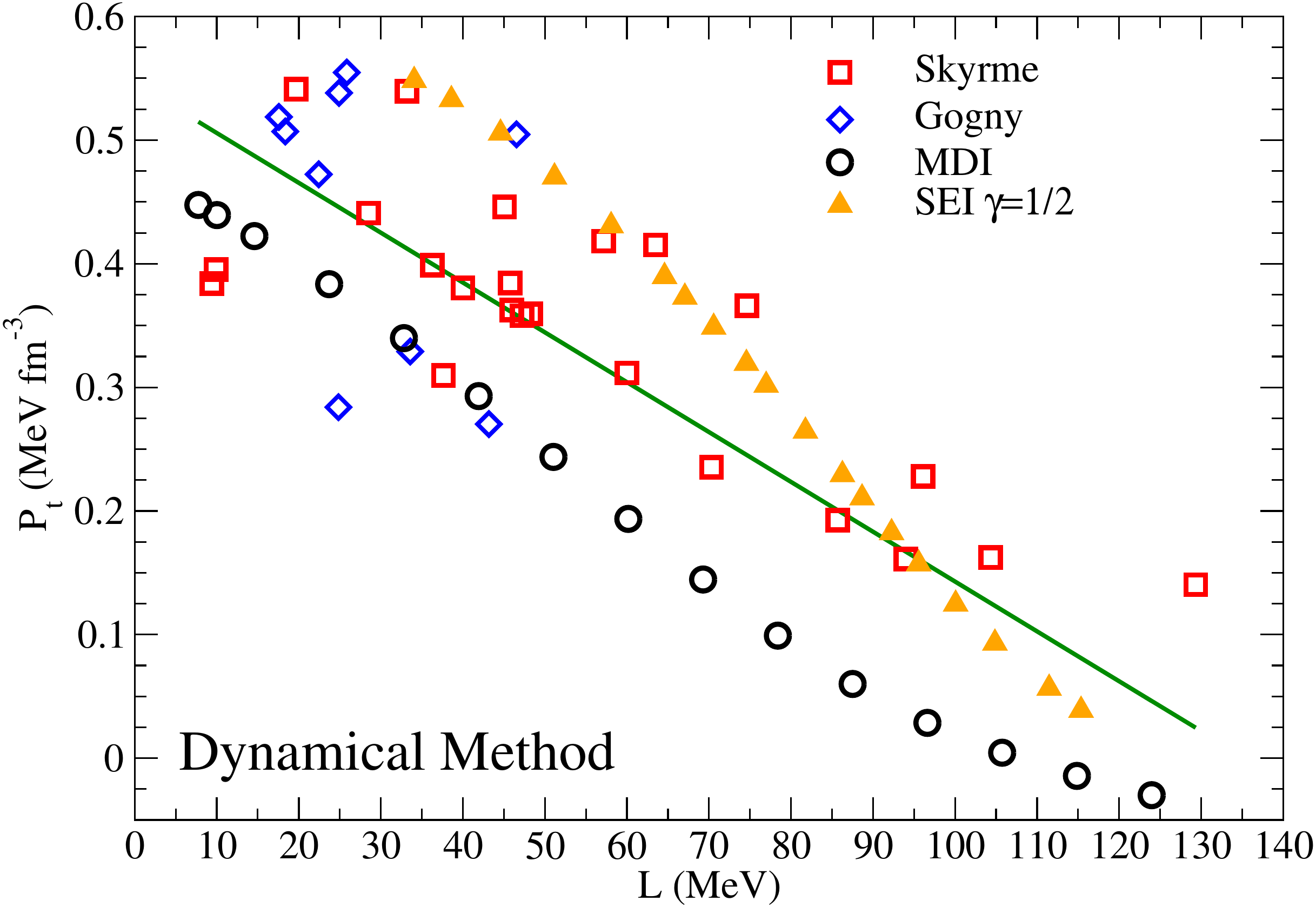}
\caption{Transition density as a function of the slope of the symmetry energy calculated using the dynamical method for 
a set of Skyrme, Gogny, MDI and SEI interactions.}\label{fig:Ptall}
\end{figure}

In Figs.~\ref{fig:rhotPtGogny}, \ref{fig:rhotPtMDI} and \ref{fig:rhotPtSEI} we display, as a 
function of the slope of the symmetry energy $L$,
the transition density (upper panels) and the transition pressure (lower panels) calculated with the 
thermodynamical ($V_\mathrm{ther}$) and the dynamical ($V_\mathrm{dyn}$) methods for three different types of finite-range interactions. 
The expression (\ref{eq14}) with the complete momentum dependence of the dynamical potential has been used for the dynamical calculations.
In the study of the transition properties we have employed several Gogny interactions \cite{decharge80,Sellahewa14,gonzalez17,gonzalez18} (Fig.~\ref{fig:rhotPtGogny}), a 
family of MDI forces \cite{das03,xu10b}, with $L$ from $7.8$ to $124.0$ MeV (corresponding to
values of the $x$ parameter ranging from
$1.15$ to $-1.4$) (Fig.~\ref{fig:rhotPtMDI}), and a family of SEI interactions \cite{behera98} with $\gamma=1/2$ and different 
values of the slope parameter $L$ (Fig.~\ref{fig:rhotPtSEI}). 
Notice that the Gogny interactions have different nuclear matter saturation properties, 
whereas the parameterizations of the MDI or SEI families displayed in Figs.~\ref{fig:rhotPtMDI} and \ref{fig:rhotPtSEI} have the
same nuclear matter saturation properties within each family.
In particular, all parametrizations of the MDI family considered here have the same nuclear matter incompressibility  
$K_0=212.6$ MeV, whereas the SEI parameterizations with $\gamma=1/2$ have $K_0= 237.5$ MeV.
Consistently with previous investigations 
(see e.g. \cite{xu09a, xu10b} and references therein) we find that for a given parameterization, the transition density 
obtained with the dynamical method is 
smaller than the prediction of the thermodynamical approach, as the surface and Coulomb contributions tend to 
further stabilize the uniform matter in the core against formation of clusters. The relative differences between the 
transition densities obtained with the thermodynamical and dynamical approaches are found to vary within the ranges
$8-14 \%$ for Gogny interactions, 
$15-30 \%$ for MDI, and $10-25 \%$ for SEI $\gamma=1/2$.
For the MDI interactions, the trends are comparable to those found in Ref.~\cite{xu09a} 
where the finite-size effects in the dynamical calculation
were taken into account phenomenologically through assumed constant values for the $D_{nn}$, $D_{pp}$ and $D_{np}=D_{pn}$ coefficients.
From the results of the transition pressure in Figs.~\ref{fig:rhotPtGogny}--\ref{fig:rhotPtSEI}, 
it is evident again that the thermodynamical 
method predicts larger values than its dynamical counterpart, which is also consistent 
with the findings in earlier works (\cite{xu09a, xu10b} and references therein).
It is to be observed that we have performed the dynamical calculations of the transition properties in two different 
ways. On the one hand, we have considered the contributions to Eqs.~(\ref{eq13}) from the direct energy only 
(the results are labelled as $V_\mathrm{dyn}$ direct in Figs.~\ref{fig:rhotPtGogny}--\ref{fig:rhotPtSEI}).
Then, we have used the complete expression of Eqs.~(\ref{eq13}) (the results are labelled as $V_\mathrm{dyn}$ direct$+ \hbar^2$ in Figs.~\ref{fig:rhotPtGogny}--\ref{fig:rhotPtSEI}).
As with our findings for the dynamical potential in the previous subsection, we see that for the three types of forces the finite-range 
effects on $\rho_t$ and $P_t$ coming from the $\hbar^2$ corrections of the exchange and kinetic energies are almost negligible
compared to the effects due to the direct energy.

Figures \ref{fig:rhotPtGogny}--\ref{fig:rhotPtSEI} also provide information about the 
dependence of the transition properties with the slope of the symmetry energy. 
We can see that within the MDI and SEI families the transition density and pressure
show a clear, nearly linear decreasing trend as a function of the slope parameter $L$. In contrast, the results
of the several Gogny forces in Fig.~\ref{fig:rhotPtGogny} show a weak trend with $L$ for $\rho_t$ and almost no trend with $L$ for $P_t$.
A more global analysis of the eventual dependence of the transition properties with the slope of 
the symmetry energy is displayed in Figs.~\ref{fig:rhotall} and \ref{fig:Ptall}. These figures include not only the results for the core-crust transition 
calculated using the considered sets of finite range interactions (Gogny, MDI and SEI of $\gamma=1/2$ interactions) but also the predictions
of $23$ zero-range Skyrme forces covering the range of $L$ between 0 and 140 MeV.
Using this 
large set of nuclear models of different nature, it can be seen that the 
decreasing trend of the transition density and pressure with rise in the slope $L$ of the symmetry energy
is a general feature.
However, the  correlation of the transition properties with the slope parameter $L$ using all interactions is
weaker than within a family of parametrizations where the saturation properties do not change. 
The correlation in the case of the transition density is found to be slightly better than in the case of
the transition pressure. The reason for the weaker correlation obtained in the
different Gogny sets (Fig.~\ref{fig:rhotPtGogny}) and Skyrme sets (Figs.~\ref{fig:rhotall} and \ref{fig:Ptall}) 
is attributed to the fact that these sets have different nuclear matter saturation properties apart from the $L$ values.

Let us mention that we have also found that in the models used in this work the correlations between the transition density and pressure with the curvature of 
the symmetry energy, defined as
\begin{equation}
 K_\mathrm{sym} = \left. 9 \rho_\mathrm{sat}^2 \frac{\partial^2 E_\mathrm{sym} (\rho)}{\partial \rho^2}\right|_{\rho_\mathrm{sat}} ,
\end{equation}
are similar in quality to those found with the slope parameter $L$, in agreement with earlier literature \cite{xu09a, Carreau19}.
\begin{figure}[t]
\centering
\includegraphics[clip=true, width=\columnwidth]{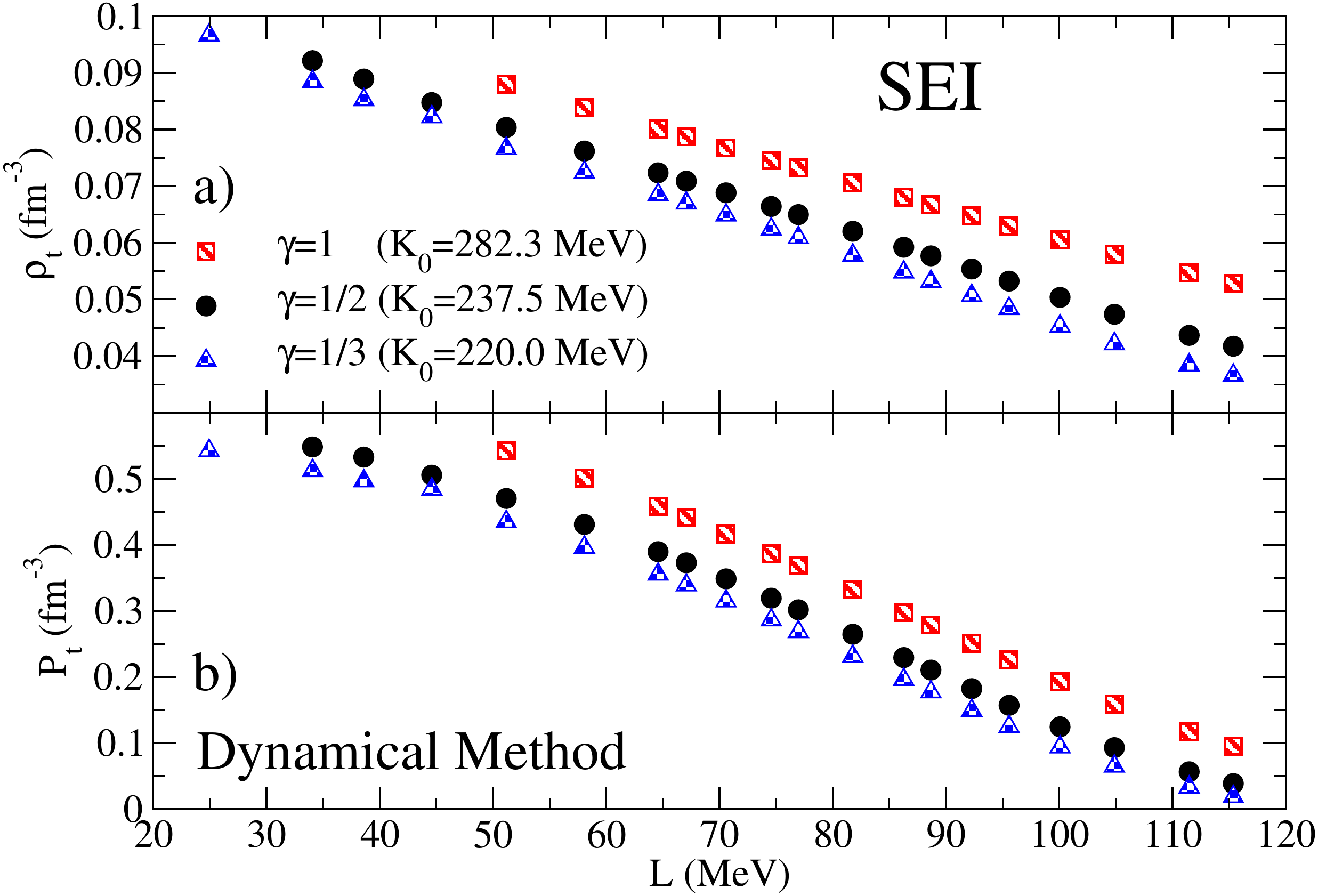}
\caption{Transition density (panel a) and transition pressure (panel b) as a function of the slope of the symmetry energy,
calculated using the dynamical method for three SEI families with different nuclear matter incompressibilities.}\label{fig:SEIK0}
\end{figure}

In order to test the impact of the nuclear matter incompressibility on the correlations
between the core-crust transition properties and the slope of the symmetry energy 
for a given type of finite-range interactions, we plot 
in Fig.~\ref{fig:SEIK0} the transition density (panel a) and the transition pressure (panel b) against the $L$ parameter for 
SEI interactions of $\gamma=1/3$, $\gamma=1/2$ and $\gamma=1$, which correspond to nuclear matter incompressibilities $K_0$ of $220.0$ MeV,
$237.5$ MeV and $282.3$ MeV, respectively, covering an extended 
range of $K_0$ values.
From this figure we observe a high correlation with $L$ for the different sets of a given SEI force having the same incompressibility
and differing only in the $L$ values. However, the comparison of the results
obtained with the SEI forces of different incompressibility shows that higher transition density and pressure are predicted for the 
force sets with higher incompressibility.
This also demonstrates, through the example of the nuclear matter incompressibility, a dependence of the features of 
the core-crust transition with the nuclear matter saturation properties of the force,
on top of the dependence on the stiffness of the symmetry energy.

\subsection{Crustal properties of the neutron star}
\begin{figure}[b]
\centering
\includegraphics[clip=true, width=\columnwidth]{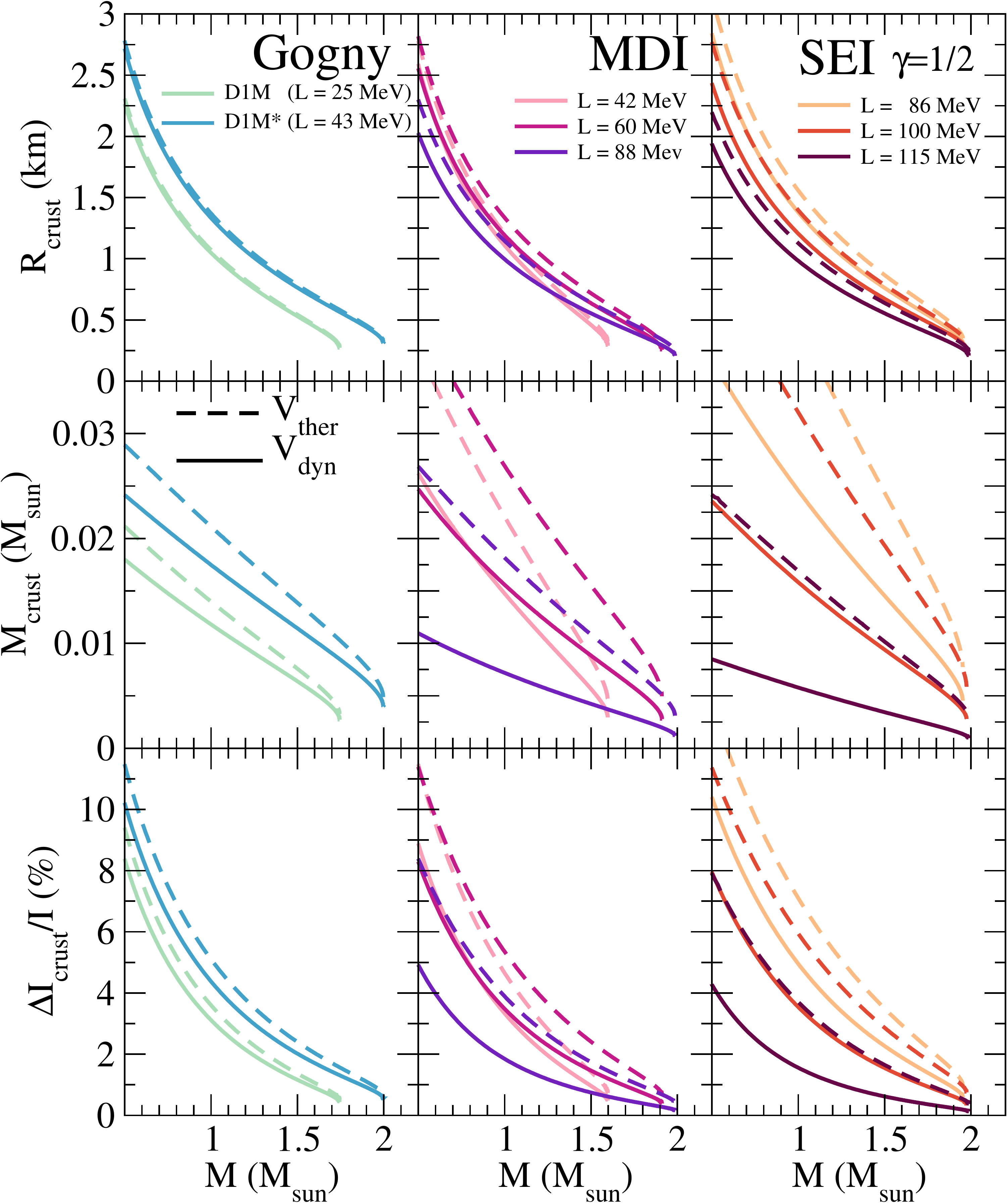}
\caption{Neutron star crustal thickness (panels a), crustal mass (panels b) and crustal fraction of the moment of inertia (panels c)
against the total mass of the neutron star for some Gogny (left), MDI (center) and SEI (right) interactions. The core-crust 
transition has been determined using the thermodynamical potential (dashed lines) and the dynamical potential 
using all direct and $\hbar^2$ contributions 
(solid lines).}\label{fig:Crust}
\end{figure}
Once we have obtained the location of the transition density and pressure, we can study
how the differences between the results of the thermodynamical and dynamical methods affect the crustal properties of neutron stars, namely
the crustal mass, crustal thickness and crustal fraction of the moment of inertia.
For that purpose, we obtain the mass and radius of the NS core by solving the TOV equations from the center of the star
to the transition point. We describe the core using the EOS of $\beta$-stable $npe \mu$ matter predicted by the finite-range forces. 
To obtain the properties of the NS crust, we use the method proposed by Zdunik, Fortin and Haensel \cite{Zdunik17},
in which only the transition density and transition pressure are needed, but 
not the explicit knowledge of the EOS of the crust.
In the work of Zdunik et al., the crustal thickness of a neutron star of mass $M$ is given by 
\begin{equation}
 R_\mathrm{crust} = \Phi R_\mathrm{core} \frac{1-2 G M / R_\mathrm{core} c^2}{1- \Phi \left( 1- 2 G M /R_\mathrm{core} c^2\right)},
\end{equation}
with   
\begin{equation}
 \Phi \equiv \frac{(\alpha-1) R_\mathrm{core} c^2}{2 G M}
\end{equation}
and 
\begin{equation}
 \alpha = \left( \frac{\mu_t}{\mu_0}\right)^2,
\end{equation}
where $\mu_t = (P_t + {\cal H}_t c^2)/\rho_t$ is the chemical potential at the core-crust transition, 
$\mu_0$ is the chemical potential at the surface of the neutron star and $R_\mathrm{core}$ is
the thickness of the core. The crustal mass is obtained as 
\begin{equation}
 M_\mathrm{crust} = \frac{4 \pi P_t R_\mathrm{core}^4}{G M_\mathrm{core}} \left(1- \frac{2 G M_\mathrm{core}}{R_\mathrm{core}c^2} \right),
\end{equation}
where $M_\mathrm{core}$ is the mass of the NS core. The total mass of the NS is $M=M_\mathrm{crust}+M_\mathrm{core}$.
This approximate approach predicts the radius and mass of the crust with
accuracy better than $\sim 1\%$ in $R_\mathrm{crust}$ and $\sim 5\%$ in $M_\mathrm{crust}$ for typical
neutron star masses \cite{Zdunik17}, while one circumvents the complex problem of
computing the EOS of the crustal matter.

Finally, to compute the crustal fraction of the moment of inertia we use the approximation given by \cite{Lattimer00, Lattimer01, Lattimer07},
which allows one to express this quantity as
\begin{eqnarray}
 \frac{\Delta I_\mathrm{crust}}{I} &=& \frac{28 \pi P_t R^3}{3 M c^2} \frac{\left(1-1.67 \xi-0.6 \xi^2 \right)}{\xi} \nonumber \\
 &&\times \left[ 1+ \frac{2 P_t \left( 1+ 5 \xi -14 \xi^2\right)}{\rho_t m c^2 \xi^2}\right]^{-1}, 
\end{eqnarray}
where $\xi= \frac{GM}{R c^2}$ is the compactness of the star, and $m$ is the baryon mass. 

We plot in Fig.~\ref{fig:Crust} the crustal thickness, crustal mass and crustal fraction of the moment of inertia against the total mass of 
the neutron star. As representative examples, we show the results provided by the D1M and D1M$^{*}$ Gogny interactions, 
three MDI and three SEI forces with different $L$ values. 
All considered forces in this figure, except from the Gogny D1M and the MDI force with $L=42$ MeV,
predict neutron stars of mass about $2 M_\odot$, in agreement with astronomical observations \cite{Demorest10, Antoniadis13}.
For each interaction we have obtained the crustal properties using the transition point given by the thermodynamical and the dynamical approaches.
In general, the global behaviour of these properties is similar for all three types of finite-range forces. 
This is true for both the methods of calculations, where the thermodynamical model gives higher values than its dynamical counterpart. 
The influence of the transition point is smaller on the crustal radius, and has a larger impact in 
the calculation of the crustal mass and crustal fraction of the moment of inertia. 
The differences between the predictions using the core-crust transition found with the thermodynamical 
or with the dynamical methods are larger for interactions with a larger value of $L$. 
These differences are very prominent in the typical NS mass region and could influence 
the properties where the crust has an important role, such as pulsar glitches~\cite{Link1999,Fattoyev:2010tb,Chamel2013,PRC90Piekarewicz2014,Newton2015, gonzalez17}.

\section{Summary and Conclusions}\label{summary}
In this work we have analyzed the core-crust transition in neutron stars estimated with the dynamical method
using several finite-range interactions of Gogny, MDI and SEI types. This study enlarges previous studies available
in the literature, which were basically performed, to our knowledge, in the case of the MDI forces. 
We follow closely the original method developed by Baym, Bethe and Pethick to derive the energy curvature matrix in momentum space.
The contribution coming from the direct energy is obtained through the
expansion of its finite-range form factors in terms of distributions, which allows one to write the direct contribution beyond
the so-called long-wavelength limit (expansion till $k^2$-order). Moreover, the Extended Thomas-Fermi expansion of the density matrix
is used to write the kinetic and exchange energies as the sum of a bulk term plus a $\hbar^2$ correction. This $\hbar^2$
term can be written as a linear combination of the square of the gradients of the neutron and proton distributions, with density-dependent coefficients,
which in turn provide a $k^2$-dependence in the energy curvature matrix.

We find that the effects of the finite-range part of the nuclear interaction on the energy curvature matrix mainly
arises from the direct part of the energy, where the $\hbar^2$ 
contributions from kinetic and exchange part of finite range interaction can be considered as small correction.

We have analyzed the global behaviour of the core-crust transition density and pressure as a function of the slope of
the symmetry energy at saturation for a large set of non-relativistic nuclear models, which include finite-range
interactions as well as Skyrme forces. 
The results for MDI sets found under the present formulation are in good agreement with the earlier ones 
reported by Xu and Ko, where they use a different
density matrix expansion to deal with the exchange energy and adopted the Vlasov equation method to obtain the curvature
energy matrix.
We also observe that within the MDI and SEI families of interactions, 
the transition density and pressure are highly correlated with the slope of the symmetry
energy at saturation. However, when nuclear models with different saturation properties are considered, these correlations are deteriorated,
in particular the one related to the transition pressure. 

Finally, we have studied the crustal properties of neutron stars, such as the crustal thickness, crustal mass and crustal 
fraction of moment of inertia, by solving the TOV equations from the center to the point corresponding to the transition 
density obtained from the dynamical as well as in the thermodynamical approach. 
It is found that the transition point determined with the thermodynamical and dynamical methods has a relevant impact on the 
considered crustal properties, in particular for models with stiff symmetry energy.
These crustal 
properties play a crucial role in the prediction of several observed phenomena, like glitches, r-mode oscillation, etc.
Therefore, the core-crust transition density needs to be ascertained as precisely as possible by 
taking into account the associated physical conditions in order to have a realistic estimation of the observed phenomena.

\section*{Acknowledgement}
C.G, M.C. and X.V. acknowledge support from Grant FIS2017-87534-P from MINECO and FEDER,
and Project MDM-2014-0369 of ICCUB (Unidad de Excelencia Mar\'{\i}a de Maeztu) from MINECO.
C.G. also acknowledges Grant BES-2015-074210 from MINECO. 
One author (TRR) thanks the Departament de  F\'isica Qu\`antica i Astrof\'isica, University of Barcelona, Spain for hospitality during the visit. 
\appendix
\begin{widetext}
\section{}\label{ApA}
The ETF approximation to the HF energy for non-local potentials
consists of replacing the quantal density matrix by the semiclassical one that contains, in
addition to the bulk (Slater) term, corrective terms depending on the second-order derivatives
of the proton and neutron densities, which account for contributions up to $\hbar^2$-order. 
The angle-averaged semiclassical ETF density matrix, derived in Refs.~\cite{centelles98,gridnev98,soubbotin00}, 
for each kind of nucleon reads 
\begin{eqnarray}
{\tilde \rho}({\bf R},s) &=&=\rho_0 ({\bf R},s) +\rho_2 ({\bf R},s) =  \frac{3j_1(k_F s)}{k_f s}\rho 
+ \frac{s^2}{216}\left\{\left[\left(9 - 2k_F\frac{f_k}{f}
- 2k_F^2\frac{f_{kk}}{f} +  k_F^2\frac{f_k^2}{f}\right)\frac{j_1(k_F s)}{k_F s} -4 j_0(k_F s)\right]
\frac{({\bf \nabla}\rho)^2}{\rho}\right. \nonumber \\
&-&\left.\left[\left(18 +  6k_F\frac{f_k}{f}\right) \frac{j_1(k_F s)}{k_F s} - 3j_0(k_F s)\right]\Delta \rho 
-\left[18\rho \frac{\Delta f}{f} + \left(18 - 6k_F \frac{f_k}{f}\right)\frac{{\bf \nabla}\rho \cdot {\bf \nabla} f}{f}\right.\right.\nonumber \\
&+& \left.\left.12 k_F \frac{{\bf \nabla}\rho\cdot {\bf \nabla} f_k}{f} - 9\rho \frac{({\bf \nabla} f)^2}{f} \right]
\frac{j_1(k_F s)}{k_F s} \right\}, \nonumber \\
\label{eqA1}
\end{eqnarray}
where $\rho ({\bf R})$ is the local density, $k_F=(3 \pi^2 \rho({\bf R}))^{1/3}$ is the corresponding Fermi momentum for each type of nucleon, and $j_1(k_F s)$
is the $l=1$ spherical Bessel function. In Eq.~(\ref{eqA1}) $f= \left.f({\bf R},k)\right|_{k=k_F}$ is the inverse of the position and momentum
dependent effective mass, defined for each type of nucleon as
\begin{equation}
f({\bf R},k) = \frac{m}{m^{*}({\bf R},k)}= 1 + \frac{m}{\hbar^2k}\frac{\partial V^{F}_{0}({\bf R},k)}{\partial k}.
\label{eqA2}
\end{equation} 
computed at the $k=k_{F}$ (see below), and $f_{k}= \left.f_{k}({\bf R},k)\right|_{k=k_F}$ and 
$f_{kk}= \left.f_{kk}({\bf R},k)\right|_{k=k_F}$ are its first and second
derivatives with respect to $k$.
In Eq.~(\ref{eqA2}), $V^{F}_{0}({\bf R},k)$ is the Wigner transform of the exchange potential 
(\ref{eq4}) given by
\begin{eqnarray}\label{eqA3} 
V^{F}_0({\bf R},k) = \int d{\bf s} V^{F}_0({\bf R},s)e^{-i{\bf k}\cdot{\bf s}} 
= \int d{\bf s} V_0^{F} ({\bf R},s) j_0 (k,s),
\end{eqnarray}
which leads to Eq.~(\ref{eqA9}) of the main text.
Notice that due to the structure of the exchange potential (\ref{eqA9}), the space dependence of the effective mass for each 
kind of nucleon is through the Fermi momenta of both type of nucleon, neutrons and protons,
 i.e. $ f_q=f_q(k,k_{Fn}({\bf R}),k_{Fp}({\bf R}))$ ($q=n,p$).
When the inverse 
effective mass and its derivatives with respect to $k$ are used in (\ref{eqA1}), an additional space
dependence arises from the replacement of the momentum $k$ by the local Fermi momentum
$k_F({\bf R})$.
 
Using the DM (\ref{eqA1}) the explicit form of the semiclassical kinetic energy at the ETF level for either neutrons or protons 
can be written as:
\begin{eqnarray}
\tau_{ETF}({\bf R}) &=& \left.\left(\frac{1}{4}\Delta_R - \Delta_s\right) {\tilde \rho}({\bf R},s)\right|_{s=0} =
\tau_{0} + \tau_{2},
\label{eqA4}
\end{eqnarray}
which consists of the well-known TF term
\begin{equation}
 \tau_{0}=\frac{3}{5}k_{F}^2\rho,
\end{equation}
plus the $\hbar^2$ contribution
\begin{eqnarray}
\tau_{2}({\bf R}) &=& 
\frac{1}{36}\frac{({\bf \nabla} \rho)^2}{\rho} \left[ 1 + \frac{2}{3}k_F \frac{f_k}{f} +
\frac{2}{3}k_F^2 \frac{f_{kk}}{f}- \frac{1}{3}k_F^2 \frac{f_k^2}{f^2} \right] + 
\frac{1}{12}\Delta \rho \left[4 + \frac{2}{3}k_F \frac{f_k}{f} \right]\nonumber \\
&+&  \frac{1}{6}\rho \frac{\Delta f}{f}+ \frac{1}{6}\frac{{\bf \nabla}\rho \cdot {\bf \nabla}f}{f}
\left[1 - \frac{1}{3}k_F \frac{f_k}{f}\right] + \frac{1}{9}\frac{{\bf \nabla}\rho \cdot {\bf \nabla}f_k}{f}
- \frac{1}{12}\rho \frac{({\bf \nabla}f)^2}{f^2}.
\label{eqA41}
\end{eqnarray}
This $\hbar^2$ contribution reduces to the standard $\hbar^2$
expression for local forces \cite{brack85} if the effective mass depends only on the 
position and not on the momentum.

The direct energy is easily obtained using the diagonal part of the DM, which at ETF level reduces to the local density
$\rho$. The exchange energy density, which is local within the ETF approximation, is obtained from the exchange potential (\ref{eqA3}) and
is given by
\begin{eqnarray}
{\cal H}_{exch}({\bf R}) = 
\frac{1}{2 }\int d{\bf s} \rho_0({\bf R},s) V_0^F ({\bf R},s) +
\int d{\bf s} \rho_2({\bf R},s) V^{F}_{0}({\bf R},s),
\label{eqA5}
\end{eqnarray}
from where, and after some algebra explained in detail in Ref.~\cite{soubbotin00}, one can recast the $\hbar^2$ 
contribution to the exchange energy for each kind of nucleon as
\begin{equation}
{\cal H}_{exch,2}({\bf R}) = 
 \frac{\hbar^2}{2M} \left[(f-1)\left(\tau_{ETF}-\frac{3}{5}k_F^2 \rho - \frac{1}{4}\Delta \rho \right) 
+ k_F f_k \left(\frac{1}{27}\frac{({\bf \nabla} \rho)^2}{\rho} - \frac{1}{36}\Delta \rho\right)\right].
\label{eqA6}
\end{equation}
Notice that $\tau_{ETF}-\frac{3}{5}k_{F}^2\rho=\tau_{2}$ and that 
$\Delta \rho$ vanishes under integral sign if spherical symmetry in coordinate space is assumed. Therefore
the sum of the total kinetic energy and the $\hbar^2$ contribution to exchange energy can be written as the pure 
TF kinetic part, which contributes to the bulk energy, plus an additional $\hbar^2$ term, which depends 
on the local Fermi momentum $k_F$ and on second-order derivatives of the nuclear density:
\begin{equation}
\int d{\bf R}\left[{\cal H}_{kin}({\bf R}) +  {\cal H}_{exch,2}({\bf R})\right] = 
\frac{\hbar^2}{2M}\int d{\bf R}\left\{\tau_{0} +  
 \left[f\tau_{2} - \frac{1}{4}f \Delta \rho 
+ k_F f_k \left(\frac{1}{27}
\frac{({\bf \nabla} \rho)^2}{\rho} - \frac{1}{36}\Delta \rho\right)\right]\right\}.
\label{eqA7}
\end{equation}
By performing a suitable partial integration of the Laplacian of the density in 
Eqs.~(\ref{eqA41}) and (\ref{eqA7}), one can recast the $\hbar^2$ contribution to (\ref{eqA7}) 
as a function of the density multiplied by the square gradient of the density as 
we have pointed out in the main text (see Eq.~(\ref{eq5a})). 

The full $\hbar^2$ contribution to the total energy corresponding to the finite-range central interaction 
(\ref{eqVfin}), given by Eq.~(\ref{eqA7}) for each kind of nucleon, can be written, after partial integration, as:
\begin{equation}
E_{\hbar^2} = \int d{\bf R}\bigg[B_{nn}(\rho_{n},\rho_{p})\big({\bf \nabla}\rho_n\big)^2
+ B_{pp}(\rho_{n},\rho_{p})\big({\bf \nabla}\rho_p\big)^2
+ 2 B_{np}(\rho_{n},\rho_{p}){\bf \nabla}\rho_n\cdot{\bf \nabla}\rho_p\bigg],
\label{eqA10}
\end{equation}
where the like coefficients of the gradients of the densities are
\begin{eqnarray}
B_{nn}(\rho_n,\rho_p) &=& \frac{\hbar^2}{2M} \frac{1}{108}\left\{\left[3 f_n + k_{Fn}(2f_{nk} - 3 f_{nk_{Fn}})
+  k^2_{Fn}(5f_{nkk} + 3 f_{nkk_{Fn}})  \right. \right.
\nonumber \\
&-& \left. \left. k^2_{Fn}\frac{(2f_{nk} +  f_{nk_{Fn}})^2}{f_n}\right]\frac{1}{\rho_n}
- \frac{\rho_p}{\rho_n^2}k^2_{Fn}\frac{f^2_{pk_{Fn}}}{f_p}\right\}
\label{eqA11}
\end{eqnarray}
and a similar expression for $B_{pp}(\rho_n,\rho_p)$ obtained by exchanging $n$ by $p$ in Eq.~(\ref{eqA11}). The 
unlike coefficient  $B_{np}(\rho_n,\rho_p)$ of Eq.~(\ref{eqA10}) reads:
\begin{eqnarray}
B_{np}(\rho_n,\rho_p)&=& B_{pn}(\rho_p,\rho_n)= - \frac{\hbar^2}{2M} \frac{1}{316}\left\{\left[3 k_{Fp}f_{nk_{Fp}}
- 3 k_{Fn} k_{Fp}f_{nkk_{Fp}} +  \frac{2 k_{Fn} k_{Fp}(2f_{nk} + f_{nk_{Fn}})f_{nk_{Fp}}}{f_n}\right]
\frac{1}{\rho_p}\right.
\nonumber \\ 
&+& \left.\left[3 k_{Fn}f_{pk_{Fn}} - 3 k_{Fp} k_{Fn}f_{pkk_{Fn}} +  
\frac{2 k_{Fp} k_{Fn}(2f_{pk} + f_{pk_{Fp}})f_{pk_{Fn}}}{f_p}\right]\frac{1}{\rho_n}\right\}.
\label{eqA12}
\end{eqnarray}
As mentioned before, all derivatives of the neutron (proton) inverse effective mass $f_q (k, k_{Fq}, k_{Fq'})$ 
with respect to the momentum $k$, $f_{qk}( k_{Fq}, k_{Fq'})$, are 
evaluated at the neutron (proton) Fermi momentum $k_{Fq}$, i.e.
$f_{qk}  (k_{Fq}, k_{Fq'}) =\left.\frac{\partial f_q (k, k_{Fq}, k_{Fq'})}{\partial k}\right|_{k=k_{Fq}}$,
$f_{qkk}  (k_{Fq}, k_{Fq'}) =\left.\frac{\partial^2 f_q (k, k_{Fq}, k_{Fq'})}{\partial k^2}\right|_{k=k_{Fq}}$,
$f_{qkk_{Fq'}}  (k_{Fq}, k_{Fq'}) =\left.\frac{\partial^2 f_q (k, k_{Fq}, k_{Fq'})}{\partial k \partial k_{Fq'}}\right|_{k=k_{Fq}}$, etc.

\section{}\label{AppB}
In this Appendix we derive the direct term in Eq.~(\ref{eq9}) due to the fluctuating density.
Let us first obtain the gradient expansion of the direct energy coming from the finite-range part of the force. For the sake of 
simplicity we consider a single Wigner term. The result for the case including spin and isospin exchange operators can be obtained analogously.
In the case of a Wigner term we have 
\begin{equation}\label{eq:Edir}
 E_{dir} = \frac{1}{2}\int d{\bf R} d{\bf s} \rho ({\bf R}) \rho({\bf R}-{\bf s}) v ({\bf s}).
\end{equation}
Following the procedure of 
Ref.~\cite{durand93}, a central finite-range interaction can be expanded in a series of distributions as follows:  
\begin{equation}
v({s}) = \sum_{n=0}^{\infty} c_{2n}\nabla^{2n} \delta ({\bf s}),
\label{eqB2}
\end{equation}
where the coefficients $c_{2n}$ are chosen in such a way 
that the expansion (\ref{eqB2}) gives the same moments of the interaction $v$($s$). 
This implies that \cite{durand93}
\begin{equation}
c_{2n} = \frac{1}{(2n+1)!} \int d{\bf s} s^{2n} v({s}),
\label{eqB3}
\end{equation}
which allows one to determine the values of the coefficients $c_{2n}$ for any value of $n$.
Using this expansion, the direct energy (\ref{eq:Edir}) can be written as 
\begin{eqnarray}
E_{dir}= \sum_{n=0}^{\infty} \frac{c_{2n}}{2} \int d{\bf R} d{\bf s} \rho({\bf R}) 
\rho({\bf R}- {\bf s}) \nabla^{2n} \delta({\bf s})
=\sum_{n=0}^{\infty}  \frac{c_{2n}}{2} \int d{\bf R} \rho({\bf R})\nabla^{2n} \rho({\bf R}).
\label{eqB4}
\end{eqnarray}      
Expanding now the density $\rho({\bf R})$ in its uniform and varying contributions,
the direct energy due to the fluctuating part of the density
becomes:
\begin{equation}
\delta E_{dir} =  \sum_{n=0}^{\infty} \frac{c_{2n}}{2} 
\int d{\bf R} \delta \rho({\bf R})\nabla^{2n} \delta \rho({\bf R}).
\label{eqB5}
\end{equation}
Notice that the splitting of the density (\ref{eq6}) also provides a contribution to the
non-fluctuating energy $E_0({\rho_0})$ through the constant density $\rho_0$.
Linear terms in $\delta \rho ({\bf R})$ do not contribute 
to the direct energy by the reasons discussed in the main text (Section \ref{SecTransition}).
Proceeding as in the main text to transform integrals in coordinate space into
integrals in momentum space (see Eqs.~(\ref{eq8}) and (\ref{eqB1})) after some algebra the fluctuating correction to 
the direct energy can be written as follows:
\begin{equation}
\delta E_{dir} = \frac{1}{2} \int \frac{d{\bf k}}{(2\pi)^3} \delta n({\bf k})\delta  n^*({\bf k}) 
{\cal F}(k),
\label{eqB6}
\end{equation}
where 
\begin{equation}
 {\cal F}(k)= \sum_{n=0}^{\infty} c_{2n}k^{2n}
\end{equation}
is a series encoding the 
response of the direct energy to the perturbation induced by the varying density.
This series is the Taylor expansion of the Fourier transform of the form factor $v(s)$. In the case of 
Gaussian ($e^{-s^2/\alpha^2}$) or Yukawian ($e^{-\mu s}/\mu s$) form factors one obtains, respectively, 
\begin{equation}
 {\cal F}(k)=\pi^{3/2} \alpha^3 e^{-\alpha^2 k^2/4}
\end{equation}
and
\begin{equation}\label{Fyuk}
 {\cal F}(k)=\frac{4\pi}{\mu(\mu^2 + k^2)}.
\end{equation}
Note that if in (\ref{Fyuk}) one takes $\mu$=0, one recovers the
result of the direct Coulomb potential \cite{baym71}.

For the nuclear direct potential the first 
 term ($c_0$) of the series for ${\cal F}(k)$, which can be written as 
${\cal F}(0)$, corresponds to the bulk contribution associated to the fluctuating density 
$\delta \rho({\bf R})$, i.e. it also contributes to $\mu_n$ and $\mu_p$ in Eq.~(\ref{eq9}). 
Then, the fluctuating correction to the direct energy in Eq.~(\ref{eq9}) is given by   
\begin{eqnarray}
 \delta E_{dir} &=& \frac{1}{2} \int \frac{d{\bf k}}{(2\pi)^3} \sum_m\left[D_{L,dir}^m\left(\delta n_n({\bf k})\delta n^*_n({\bf k})
 + \delta n_p({\bf k})\delta n^*_p({\bf k})\right) \right. \nonumber \\
&+& \left. D_{U,dir}^m\left(\delta n_n({\bf k})\delta n^*_p({\bf k})
+\delta n_p({\bf k})\delta n^*_n({\bf k})\right)
\right]({\cal F}_m(k)- {\cal F}_m(0)) 
\end{eqnarray}
Let us also point out that 
if the series ${\cal F}(k)$ is cut at first order, i.e. taking only the $n=1$ term of the series, $c_2$, one recovers 
the typical $k^2$ dependence corresponding to square gradient terms in the energy density functional. 
If this expansion up to quadratic terms in $k$ is used, the dynamical potential potential can be written 
as Eq.~(\ref{eq14a}), where the coefficient $\beta (\rho)$ reads
\begin{equation}
\beta (\rho) = \left[\sum_m D^m_{L,dir}c^m_2 + 2B_{pp} \left(\rho_n, \rho_p\right)\right] + \frac{\left(\frac{\partial \mu_p}{\partial \rho_n}\right)^2}
{\left(\frac{\partial \mu_n}{\partial \rho_n}\right)^2}\left[ \sum_m D^m_{L,dir}c^m_2 + 2B_{nn}\left(\rho_n, \rho_p\right)\right]
-2 \frac{\frac{\partial \mu_p}{\partial \rho_n}}
{\frac{\partial \mu_n}{\partial \rho_n}}\left[ \sum_m D^m_{U,dir}c^m_2 + 2B_{np}\left(\rho_n, \rho_p\right)\right],
\label{eqB8}
\end{equation}
and the $B_{qq}\left(\rho_n, \rho_p\right)$ and $B_{qq'}\left(\rho_n, \rho_p\right)$ functions have been given in 
Eqs.~(\ref{eqA11}) and (\ref{eqA12}), corresponding to the $\hbar^2$ contributions coming from the expansion of the energy
density functional. Moreover, for Gaussian form factors one has 
\begin{equation}
 c_2^m = -\frac{\pi^{3/2} \alpha_m^5}{4},
 \end{equation}
whereas for Yukawian form factors one has 
\begin{equation}
 c_2^m= -\frac{4 \pi}{\mu_m^5}.
\end{equation}

\end{widetext}
\bibliography{bibtex}
\end{document}